\documentclass[]{IEEEtranTIE}

\usepackage{graphicx}
\usepackage{cite}
\usepackage{picinpar}
\usepackage{amsmath}
\usepackage{mathrsfs}
\usepackage{url}
\usepackage[utf8]{inputenc}
\usepackage{colortbl}
\usepackage{soul}
\usepackage{multirow}
\usepackage{pifont}
\usepackage{alltt}
\usepackage[hidelinks]{hyperref}
\usepackage{enumerate}
\usepackage{breakurl}
\usepackage{epstopdf}
\usepackage{pbox}

\usepackage{float}
\usepackage{siunitx}
	\DeclareSIUnit{\var}{ var }
	\DeclareSIUnit{\voltampere}{ VA }
\usepackage[greek,english]{babel}

\usepackage{tabularx}
\usepackage{booktabs}

\usepackage{subfig}

\usepackage{mathtools}
\usepackage{amsfonts}
\usepackage{amsthm}
\usepackage{amssymb}
\usepackage{amstext}
\usepackage{bm}
\usepackage{arydshln}

\makeatletter
\renewcommand*\env@matrix[1][*\c@MaxMatrixCols c]{%
  \hskip -\arraycolsep
  \let\@ifnextchar\new@ifnextchar
  \array{#1}}
\makeatother

\usepackage{pgfplots}
	\pgfplotsset{compat=newest}
	\usepgfplotslibrary{fillbetween}
\usepackage{tikz}
\usepackage[siunitx, european, straightvoltages]{circuitikz}
\usetikzlibrary{external}

\usetikzlibrary{shapes, arrows}
\usetikzlibrary{patterns, positioning}
\usetikzlibrary{intersections}
\usetikzlibrary{decorations.markings}

\tikzset{
  declare function={
    atan3(\a,\b)=ifthenelse(atan2(0,1)==90, atan2(\a,\b), atan2(\b,\a));},
  kinky cross radius/.initial=+.125cm,
  @kinky cross/.initial=+, kinky crosses/.is choice,
  kinky crosses/left/.style={@kinky cross=-},kinky crosses/right/.style={@kinky cross=+},
  kinky cross/.style args={(#1)--(#2)}{
    to path={
      let \p{@kc@}=($(\tikztotarget)-(\tikztostart)$),
          \n{@kc@}={atan3(\p{@kc@})+180} in
      -- ($(intersection of \tikztostart--{\tikztotarget} and #1--#2)!%
             \pgfkeysvalueof{/tikz/kinky cross radius}!(\tikztostart)$)
      arc [ radius     =\pgfkeysvalueof{/tikz/kinky cross radius},
            start angle=\n{@kc@},
            delta angle=\pgfkeysvalueof{/tikz/@kinky cross}180 ]
      -- (\tikztotarget)}}}

\tikzset{->-/.style={decoration={markings,mark=at position #1 with {\arrow{latex}}},postaction={decorate}}}

\ctikzset{voltage/distance from node=.2,voltage/european label distance=1,}
\ctikzset{voltage arrow color/.initial=jkc3}

\pgfkeys{
	/pgfplots/every y tick label/.append style  =
		{ 
			/pgf/number format/.cd, fixed relative,precision=3,1000 sep={\,}
		}
	}

\tikzset{
    o/.style={
        decoration={
            markings,
            mark={
                at position 0
                with {
                    \fill circle [radius=#1];
                }
            }
        },
        postaction=decorate
    },
    o/.default=2pt
}

\tikzset{
    dline/.style={
        double,
        double distance=2pt,
        thick
    }
}

\tikzset{
  -|-/.style={
    to path={
      (\tikztostart) -| ($(\tikztostart)!#1!(\tikztotarget)$) |- (\tikztotarget)
      \tikztonodes
    }
  },
  -|-/.default=0.5,
  |-|/.style={
    to path={
      (\tikztostart) |- ($(\tikztostart)!#1!(\tikztotarget)$) -| (\tikztotarget)
      \tikztonodes
    }
  },
  |-|/.default=0.5,
}

\newcommand{\BDBlockPIC}[7][1.5]{
	\node[very thick, fill=white, draw, rectangle, minimum width=#1 cm, minimum height=#1 cm] (#2) at (#3) {};
	\draw[stealth-stealth] (#2) ++ (-0.35*#1, 0.35*#1) -- ++ (0,-0.7*#1) -- ++ (0.7*#1,0);
	\draw[#4, very thick] (#2) ++ (-0.35*#1, -0.35*#1) -- ++ (0, 0.35*#1) -- ++ (0.65*#1,0.3*#1);
	
	\node[anchor=south west] at ($(#2) + (-0.65*#1,0.5*#1)$) {#5};
	\node[anchor=south east] at ($(#2) + (0.65*#1,0.5*#1)$) {#6};
	\node[anchor=south] at ($(#2) + (0, 0.8*#1)$) {#7};
}

\newcommand{\BDBlockSat}[5][1.5]{
	\node[thick, fill=white, draw, rectangle, minimum width=#1 cm, minimum height=#1 cm] (#2) at (#3) {};
	\node[thick, fill=white, draw, rectangle, minimum width=.9*#1 cm, minimum height=.9*#1 cm] at (#2) at (#3) {};
	\draw[-stealth] (#2) ++ (0, -0.4*#1) -- ++ (0,0.8*#1);
	\draw[-stealth] (#2) ++ (-0.4*#1, 0) -- ++ (0.8*#1,0);
	
	\draw[#4, very thick] (#2) ++ (-0.4*#1, -0.2*#1) -- ++ (0.2*#1, 0) -- ++ (0.4*#1,0.4*#1) -- ++ (0.2*#1,0);
	
	\node[anchor=south] at ($(#2) + (0, 0.65*#1)$) {#5};
}

\newcommand{\BDBlockTransform}[6][1.5]{
	\node[very thick, fill=white, draw, rectangle, minimum width=#1 cm, minimum height=#1 cm, #6] (#2) at (#3) {};
	
	\draw[] (#2.north east) -- (#2.south west);
	\node[below right] at (#2.north west) {\large #4};
	\node[above left, anchor=south east] at (#2.south east) {\large #5};
}

\newcommand{\BDBusDemux}[4][1.5]{
	\pgfmathsetmacro{\DHEIGHT}{#1}
	\pgfmathsetmacro{\DMARGIN}{#1*0.25}
	\node[fill=black, rectangle, minimum height=\DHEIGHT cm, minimum width=0.125*#1 cm, inner sep=0] (#2) at (#3) {};
		\foreach \i in {1,...,#4}
		{
			\coordinate (#2_e\i) at ($(#2.north east) + (0, \DMARGIN-\i*#1/#4)$);
			\coordinate (#2_w\i) at ($(#2.north west) + (0, \DMARGIN-\i*#1/#4)$);
		}
}

\newcommand{\BDSumCirc}[3][1.5]{
	\node[draw, circle, thick, minimum width=.2*#1 cm, inner sep=0, fill=white] (#2) at (#3) {};
}

\newcommand{\BDBlockLinear}[5][1.5]{
	\node[very thick, fill=white, draw, rectangle, minimum width=#1 cm, minimum height=#1 cm, #5] (#2) at (#3) {#4};
	
	\coordinate (#2_a) at ($(#3) + (-0.5*#1, 0.3*#1)$);	
	\coordinate (#2_b) at ($(#3) + (-0.5*#1, -0.3*#1)$);
	\coordinate (#2_c) at ($(#3) + (+0.5*#1, 0.3*#1)$);	
	\coordinate (#2_d) at ($(#3) + (+0.5*#1, -0.3*#1)$);
	\coordinate (#2_e) at ($(#3) + (-0.3*#1, +0.5*#1)$);	
	\coordinate (#2_f) at ($(#3) + (+0.3*#1, +0.5*#1)$);
	\coordinate (#2_g) at ($(#3) + (-0.3*#1, -0.5*#1)$);	
	\coordinate (#2_h) at ($(#3) + (+0.3*#1, -0.5*#1)$);
}

\newcommand{\BDBlockNonLinear}[5][1.5]{
	\node[thick, fill=white, draw, rectangle, minimum width=#1 cm, minimum height=#1 cm, #5] (#2) at (#3) {};
	\node[thick, fill=white, draw, rectangle, minimum width=.9*#1 cm, minimum height=.9*#1 cm, inner sep=0] at (#3) {#4};
	
	\coordinate (#2_a) at ($(#3) + (-0.5*#1, 0.3*#1)$);	
	\coordinate (#2_b) at ($(#3) + (-0.5*#1, -0.3*#1)$);
	\coordinate (#2_c) at ($(#3) + (+0.5*#1, 0.3*#1)$);	
	\coordinate (#2_d) at ($(#3) + (+0.5*#1, -0.3*#1)$);
	\coordinate (#2_e) at ($(#3) + (-0.3*#1, +0.5*#1)$);	
	\coordinate (#2_f) at ($(#3) + (+0.3*#1, +0.5*#1)$);
	\coordinate (#2_g) at ($(#3) + (-0.3*#1, -0.5*#1)$);	
	\coordinate (#2_h) at ($(#3) + (+0.3*#1, -0.5*#1)$);
}

\newcommand{\mv}[1]{\bm{#1}}
\newcommand{\mm}[1]{\bm{#1}}

\newcommand{\J}{\mm{J}}

\newcommand{\ov}[1]{\mm{0}_{#1}}
\newcommand{\I}[1]{\mv{I}_{#1}}

\newcommand{\symtorque}{m}

\newcommand{\symvolt}{u}
\newcommand{\symcurrent}{i}
\newcommand{\symfluxl}{\psi}
\newcommand{\symresi}{R}
\newcommand{\symindself}{L}

\newcommand{\symangvel}{\omega}
\newcommand{\symangle}{\phi}

\newcommand{\symfrequency}{f}

\newcommand{\idxstator}{\mathrm{s}}
\newcommand{\idxrotor}{\mathrm{r}}

\newcommand{\idxmech}{\mathrm{m}}

\newcommand{\idxcore}{\mathrm{c}}

\newcommand{\idxsampling}{\mathrm{S}}

\newcommand{\idxrated}{\mathrm{N}}

\newcommand{\csgabcsym}{abc}

\newcommand{\csgabc}{\csgabcsym}

\newcommand{\csgdqsym}{dq}

\newcommand{\csgdq}{\csgdqsym}
\newcommand{\csgd}{d}
\newcommand{\csgq}{q}

\newcommand{\symref}{\star}

\newcommand{\R}{\mathbb{R}}

\newcommand{\N}{\mathbb{N}}

\newcommand{\usabc}{\mv{\symvolt}_{\idxstator}^{\csgabc}}

\newcommand{\usabcsatref}{\mv{\symvolt}_{\idxstator,\mathrm{sat}}^{\csgabc\symref}}

\newcommand{\usdq}{\mv{\symvolt}_{\idxstator}^{\csgdq}}

\newcommand{\usd}{\symvolt_{\idxstator}^{\csgd}}
\newcommand{\usq}{\symvolt_{\idxstator}^{\csgq}}

\newcommand{\usdqref}{\mv{\symvolt}_{\idxstator}^{\csgdq\symref}}
\newcommand{\usdqsatref}{\mv{\symvolt}_{\idxstator,\mathrm{sat}}^{\csgdq\symref}}

\newcommand{\usdqrefPI}{\mv{\symvolt}_{\idxstator,\mathrm{pi}}^{\csgdq\symref}}
\newcommand{\usdrefPI}{\symvolt_{\idxstator,\mathrm{pi}}^{\csgd\symref}}\newcommand{\usqrefPI}{\symvolt_{\idxstator,\mathrm{pi}}^{\csgq\symref}}
\newcommand{\usdqrefFF}{\mv{\symvolt}_{\idxstator,\mathrm{ff}}^{\csgdq\symref}}
\newcommand{\usdrefFF}{\symvolt_{\idxstator,\mathrm{ff}}^{\csgd\symref}}\newcommand{\usqrefFF}{\symvolt_{\idxstator,\mathrm{ff}}^{\csgq\symref}}

\newcommand{\isplain}{\mv{\symcurrent}_{\idxstator}}
\newcommand{\isabc}{\mv{\symcurrent}_{\idxstator}^{\csgabc}}

\newcommand{\isdq}{\mv{\symcurrent}_{\idxstator}^{\csgdq}}
\newcommand{\isnom}{\symcurrent_{\idxstator,\idxrated}}
\newcommand{\isdmin}{\symcurrent_{\idxstator,\mathrm{min}}^\csgd}

\newcommand{\isd}{\symcurrent_{\idxstator}^{\csgd}}
\newcommand{\isq}{\symcurrent_{\idxstator}^{\csgq}}
\newcommand{\isdgrid}{\bar{\symcurrent}_{\idxstator}^{\csgd}}
\newcommand{\isqgrid}{\bar{\symcurrent}_{\idxstator}^{\csgq}}
\newcommand{\isdref}{\symcurrent_{\idxstator}^{\csgd^\star}}
\newcommand{\isqref}{\symcurrent_{\idxstator}^{\csgq^\star}}
\newcommand{\isdqref}{\mv{\symcurrent}_{\idxstator}^{\csgdq\symref}}

\newcommand{\imdq}{\mv{\symcurrent}_{\mathrm{m}}^{\csgdq}}
\newcommand{\ifedq}{\mv{\symcurrent}_{\idxcore}^{\csgdq}}

\newcommand{\irdq}{\mv{\symcurrent}_{\idxrotor}^{\csgdq}}

\newcommand{\psisdq}{\mv{\symfluxl}_{\idxstator}^{\csgdq}}

\newcommand{\psisq}{\symfluxl_{\idxstator}^{\csgq}}

\newcommand{\psissidq}{\mv{\symfluxl}_{\idxstator\sigma}^{\csgdq}}

\newcommand{\psimdq}{\mv{\symfluxl}_{\mathrm{m}}^{\csgdq}}

\newcommand{\psirdq}{\mv{\symfluxl}_{\idxrotor}^{\csgdq}}

\newcommand{\psird}{\symfluxl_{\idxrotor}^{\csgd}}
\newcommand{\psirq}{\symfluxl_{\idxrotor}^{\csgq}}
\newcommand{\psirdqhat}{\mv{\hat{\symfluxl}}_{\idxrotor}^{\csgdq}}
\newcommand{\psirdhat}{\hat{\symfluxl}_{\idxrotor}^{\csgd}}

\newcommand{\psirsidq}{\mv{\symfluxl}_{\idxrotor\sigma}^{\csgdq}}

\newcommand{\Rs}{\symresi_{\idxstator}}

\newcommand{\Rr}{\symresi_{\idxrotor}}

\newcommand{\Rfe}{\symresi_{\idxcore}}

\newcommand{\Lssi}{\symindself_{\idxstator\sigma}}
\newcommand{\Ls}{\symindself_{\idxstator}}
\newcommand{\Lrsi}{\symindself_{\idxrotor\sigma}}
\newcommand{\Lr}{\symindself_{\idxrotor}}
\newcommand{\Lm}{\symindself_{m}}

\newcommand{\Tr}{T_\idxrotor}

\newcommand{\pel}{p_{\mathrm{e}}}
\newcommand{\ploss}{p_{\mathrm{l}}}

\newcommand{\pme}{p_{\idxmech}}

\newcommand{\omegak}{\symangvel_{\mathrm{k}}}

\newcommand{\omegakhat}{\hat{\symangvel}_{\mathrm{k}}}
\newcommand{\omegar}{\symangvel_{\idxrotor}}

\newcommand{\phik}{\symangle_{\mathrm{k}}}

\newcommand{\phikhat}{\hat{\symangle}_{\mathrm{k}}}

\newcommand{\mme}{\symtorque_{\idxmech}}
\newcommand{\mfe}{\symtorque_{\idxcore}}
\newcommand{\mel}{\symtorque_{\mathrm{e}}}
\newcommand{\melhat}{\hat{\symtorque}_{\mathrm{e}}}
\newcommand{\mld}{\symtorque_{\mathrm{l}}}

\newcommand{\mfric}{\symtorque_{\mathrm{f}}}

\newcommand{\melnom}{\symtorque_{\mathrm{e},\idxrated}}
\newcommand{\melref}{\symtorque_{\mathrm{e}}^\symref}

\newcommand{\np}{n_\mathrm{p}}

\newcommand{\ddt}{\tfrac{\text{d}}{\mathrm{d}t}}
\newcommand{\dt}{\mathrm{d}t}

\newcommand{\omegam}{\omega_\idxmech}

\newcommand{\udc}{\symvolt_\mathrm{dc}}

\newcommand{\fs}{\symfrequency_{\idxsampling}}

\newcommand{\Thetam}{\Theta_\idxmech}

\newcommand{\eobsdq}[1][]{\mv{e}_{#1}^{\csgdq}}

\newcommand{\Kiis}{K_{\mathrm{i}}}
\newcommand{\Kpis}{K_{\mathrm{p}}}

\usepackage{xcolor}
	\definecolor{jkc3}{RGB}{90,147,103}
	\definecolor{jkc4}{RGB}{247,178,103}
	\definecolor{jkc5}{RGB}{125,46,143}
	\definecolor{jkc1}{RGB}{74,109,124}
	\definecolor{jkc2}{RGB}{219,90,66}
	\definecolor{jkc0}{RGB}{181,181,181}

	\definecolor{plco2}{RGB}{89,136,176}
	\definecolor{plco1}{RGB}{233,72,73}
	\definecolor{plco3}{RGB}{113,190,110}
	\definecolor{plco5}{RGB}{255,153,51}
	\definecolor{plco4}{RGB}{173,113,181}
	\definecolor{plco6}{RGB}{175,103,62}

	\definecolor{plc1}{RGB}{88,137,176}
	\definecolor{plc2}{RGB}{233,72,73}
	\definecolor{plc3}{RGB}{113,191,110}
	\definecolor{plc4}{RGB}{255,152,51}
	\definecolor{plc5}{RGB}{172,113,181}
	\definecolor{plc6}{RGB}{250,234,162}
	\definecolor{plc7}{RGB}{253,193,109}
	\definecolor{plc8}{RGB}{246,122,69}
	\definecolor{plc9}{RGB}{217,67,78}
	\definecolor{plc10}{RGB}{158,1,66}
	
	\definecolor{plca1}{RGB}{27, 23, 137}
	\definecolor{plca2}{RGB}{198, 59, 138}
	\definecolor{plca3}{RGB}{179, 0, 101}
	\definecolor{plca4}{RGB}{59, 192, 222}
	\definecolor{plca4b}{RGB}{33, 101, 122}
	\definecolor{plca5}{RGB}{18, 137, 83}
	\definecolor{plca5b}{RGB}{9, 69, 42}
	
\usepackage{fancyhdr}

\usepackage{empheq}
\newtheorem{remark}{Remark}

\usepackage{listings}
\lstdefinestyle{thesis}{
    backgroundcolor=\color{white},   
    commentstyle=\color{c6},
    keywordstyle=\color{c4},
    numberstyle=\tiny\color{c2},
    stringstyle=\color{c5},
    basicstyle=\footnotesize,
    breakatwhitespace=false,         
    breaklines=true,                 
    captionpos=b,                    
    keepspaces=true,                 
    numbers=left,                    
    numbersep=5pt,                  
    showspaces=false,                
    showstringspaces=false,
    showtabs=false,                  
    tabsize=2
}
\begin{document}
\title{Generic machine identification and maximum efficiency operation of induction machines}

\author{
	\vskip 1em
 	Julian Kullick${}^{\dagger}$
 	and Christoph M. Hackl${}^{\ddagger}$

	\thanks{
 		${}^{\dagger}$J. Kullick is with the research group “Control of renewable energy systems” (CRES) at the Munich School of Engineering (MSE), Technical University of Munich (TUM), Germany.

 		${}^{\ddagger}$C. M. Hackl is with the Department of Electrical Engineering and Information Technology at the Munich University of Applied Sciences (MUAS).
 		
 		Funding from the Bavarian State Ministry of Education, Science and the Arts in the frame of the project Geothermie-Allianz Bayern is gratefully acknowledged.
	}
}

\maketitle
	
\begin{abstract}
	This paper proposes an advanced machine identification method for inverter fed squirrel-cage induction machines, based on steady-state measurements in the rotor flux oriented $dq$-reference frame. The measured machine maps are used to extract maximum efficiency per torque (MEPT) look-up tables (LUTs), which guarantee the maximum achievable efficiency in every operating point. Furthermore, it is shown, that comparable results can be achieved, even without a torque sensor. The main advantage of the described method is its generality, which implicitly covers magnetic saturation, iron losses and other nonlinear effects that are typically neglected or approximated by complex models. Finally, the efficiencies of V/Hz and field-oriented control (FOC) are calculated for different speeds and load torques, allowing for quantitative assessment and comparison of both methods. 
\end{abstract}


\markboth{}%
{}

\definecolor{limegreen}{rgb}{0.2, 0.8, 0.2}
\definecolor{forestgreen}{rgb}{0.13, 0.55, 0.13}
\definecolor{greenhtml}{rgb}{0.0, 0.5, 0.0}

\section{Introduction}

\IEEEPARstart{I}\,nduction machines (IMs) are widely used in industry applications, due to their low cost, robustness and good field-weakening capabilities~\cite{buja2004, zhu2007electrical, boldea2008control}. Variable speed operation of IMs is typically realized by either scalar---e.g. V/Hz (V/f, U/f) control---or vector control methods---e.g. field-oriented control (FOC) or direct torque control (DTC). 
In open-loop V/Hz control, three-phase sinusoidal voltages with fixed amplitude-to-frequency ratio are applied to the motor terminals. The applied frequency determines the approximate mechanical speed of the machine, depending on the acting load torque. However, due to the absence of speed feedback, an error between the applied frequency and the actual speed (slip speed) results from the inherent properties of the IM. In spite of this, V/Hz control represents a low cost and reliable speed-sensorless control solution, which is sufficient in many industry applications, where dynamic performance and accurate speed tracking are not of highest priority~\cite{finch2008controlled}. In contrast, FOC provides superior dynamic performance and accurate torque control~\cite{sen1990}. By directly regulating the magnetizing ($d$) and torque producing ($q$) current components of the machine, the flux and torque  can be set independently. The major drawback of FOC is the required speed information, which is needed for the field orientation. 
However, since for vector controlled machines infinitely many combinations of $d$- and $q$-currents produce the same torque, this degree of freedom may be exploited for secondary objectives, such as efficiency optimization.

Maximum efficiency operation (also known as \emph{loss minimizing control}) of induction machines has been subject to extensive research in the past, with two main approaches having emerged: (i) offline calculation of optimum controller set points (e.g.~\cite{bojoi2013, odhano2015}), independent of the employed control system, and (ii) search-based online techniques (e.g. \cite{wasynczuk1998, uddin2008new, uddin2008development, qu2012}), often directly incorporated into the control system. 

As an extension to the classical maximum torque per ampere control (MTPA) strategy, which minimizes the copper losses in the stator windings and rotors bars, respectively, e.g. in \cite{wasynczuk1998, bojoi2013}, the maximum efficiency [or maximum efficiency per torque (MEPT)] control strategy considers also the speed dependent iron losses, which are more difficult to model and hence make up the main part of the research activity.

As an example, Qu et al. derive a steady-state iron loss model in \cite{qu2012}, which is used for online calculation of the optimum current references of a speed-sensorless PI current control system. A similar model is used by Uddin and Nam \cite{uddin2008development}, who incorporate the loss model in a nonlinear backstepping controller. A more sophisticated model is used by Pfingsten et al. \cite{pfingsten2016, pfingsten2017}, who consider transient iron losses, as well as harmonic losses. Moreover, a transient iron loss model is used by Borisevich and Schullerus~\cite{borisevich2016}, who point out the difference between energy losses calculated by steady-state and transient models, respectively.

Another important aspect to consider, is the nonlinear flux characteristic---due to magnetic saturation---and the resulting nonlinear main inductance which affect the torque production and, hence, the efficiency of the machine. Identification of the flux and inductance curves is therefore an important task, which has been treated e.g. by Wang et al.~\cite{wang2015magnetizing} and Odhano et al.~\cite{odhano2015}. The latter use parameter identification over the whole operation regime, with the results stored in LUTs and used for calculation of the stator current set points. A similar approach is used by Bojoi et al.~\cite{bojoi2013}, who, however, neglect iron losses and focus on copper losses in both, stator and rotor. 
	
	In this paper, an experimental method of extracting the maximum efficiency current references for a given torque reference [also known as \emph{maximum efficiency per torque} (MEPT)] and measured speed is presented. Moreover, the acquired machine maps allow for detailed analysis of the machine characteristics. The proposed method does not require an explicit iron loss model (as opposed to Odhano et al.) and implicitly covers the effect of magnetic saturation. The nonlinear flux and efficiency maps in the estimated $dq$-reference frame are shown and analyzed. Moreover, the extracted MEPT strategy is compared to other methods, such as constant flux (CF) and maximum torque per ampere (MTPA) control, revealing that an often neglected advantage of vector control over scalar control is its superior efficiency, even for standard CF control. To the best knowledge of the authors, a quantitative efficiency assessment of the aforementioned control methods (\emph{including} V/Hz control) has not been published before.

\section{Machine model and field-oriented control}
	In this section the generic electromechanical model of the squirrel-cage induction machine (SCIM) is stated and the FOC based torque control system is discussed (see Fig.~\ref{fig:control_system_overview}).
	
	\subsection{Machine model in arbitrarily rotating coordinates}
	The generic SCIM model in the rotating $dq$-reference frame---rotating at angular velocity $\omegak\in\R$, with (Park) transformation angle $\phik\in[0,\np 2\pi)$ and number of pole pairs $\np\in\R$---is given by the following set of equations (argument $t$ dropped for the sake of brevity)
	\begin{equation}
		\left.\begin{split}
			\usdq &= \Rs\isdq + \ddt\psisdq + \omegak\J\psisdq, \\
			\ov2   &= \Rr\irdq + \ddt\psirdq + (\omegak-\omegar)\J\psirdq,\\
			\Thetam\ddt\omegam &= \mel + \mfric + \mld, \\
			\ddt\phik &= \omegak,
		\end{split}\quad\right\}
		\label{eq:machine_model}
	\end{equation}
	with stator voltage $\usdq\in \R^2$, stator and rotor currents $\isdq, \irdq \in\R^2$, stator and rotor flux linkages $\psisdq, \psirdq\top \in \R^2$, stator and rotor resistances $\Rs, \Rr > 0$, electrical rotor speed $\omegar = \np\omegam \in \R$, mechanical rotor speed $\omegam\in\R$ and rotation matrix $\J := \begin{bsmallmatrix}0&-1\\1& 0\end{bsmallmatrix}$. Moreover, $\Thetam > 0$ is the total moment of inertia, and $\mel$, $\mfric$ and $\mld \in\R$ denote the machine, friction and load torque, respectively. Furthermore, the torque \emph{acting on the rotor} is given by
	\begin{equation}
		\mel = \underbrace{-\frac{3}{2}\np{\irdq}^\top\J\psirdq}_{\dagger}= \underbrace{\frac{3}{2}\np{\isdq}^\top\J\psisdq - \mfe}_{\ddagger},
		\label{eq:machine_torque}
	\end{equation}
	where $\mfe \in\R$ is a core loss related torque term (the effect of torque detuning due to core losses was thoroughly investigated by Levi et al. in~\cite{levi1996}). Its value and the equivalence of the torque expressions $\dagger$ and $\ddagger$ result from the power equilibrium (as will be shown in Sec.~\ref{sec:power_balance}).  
	%
	
	%

	\subsection{Field-oriented control of induction machines}
	\label{sec:control_system}
	Rotor flux orientation is used, which allows for an (almost) decoupled control of the rotor flux linkage and machine torque, respectively. An overview of the employed control system is depicted in Fig.~\ref{fig:control_system_overview}. 
	
	\begin{figure*}
		\resizebox{\linewidth}{!}{
		\input{./figures/tikz/control_system.tikz}}
		\caption{Overview of the field-oriented control system.}
		\label{fig:control_system_overview}
	\end{figure*}

	\subsubsection{Rotor flux estimation (RFE in Fig.~\ref{fig:control_system_overview})}
		Since the rotor flux linkage cannot be measured directly, it needs to be estimated instead. If---hypothetically---ideal rotor flux orientation was given, the total rotor flux linkage would be concentrated in its $d$-component, i.e.
		\begin{equation}
			\psirdq = \begin{pmatrix}\psird\\ \psirq\end{pmatrix} = \begin{pmatrix}\psird\\ 0\end{pmatrix}
			\label{eq:rotor_flux_linkage_ideal}
		\end{equation}
		 would hold. By further assuming linear flux linkage models $\psisdq = \Ls\isdq + \Lm\irdq$ and $\psirdq = \Lm\isdq + \Lr\irdq$, with stator and rotor self inductances $\Ls = \Lm+\Lssi>0$ and $\Lr = \Lm+\Lrsi >0$ and leakage inductances $\Lssi>0$ and $\Lrsi > 0$, respectively, the estimator dynamics are obtained by rearranging the rotor $d$-voltage equation in \eqref{eq:machine_model}, i.e.
		\begin{equation}
			\ddt\psirdhat = \tfrac{\Lm}{\Lr}\isd  - \tfrac{1}{\Tr} \psirdhat,
			\label{eq:rotor_flux_linkage_estimation}
		\end{equation}
		where $\psirdhat\in\R$ denotes the estimated rotor flux linkage and $\Tr = \tfrac{\Lr}{\Rr}>0$ is the rotor time constant. Moreover, since $\ddt\psirq = 0$ ideally holds, evaluating the rotor $q$-voltage yields
		\begin{equation}
			\omegak = \omegar + \frac{\Lm}{\Lr}\Rr\frac{\isq}{\psirdhat}, 
			\label{eq:omegak_estimation}
		\end{equation}
		which is valid for all $\psirdhat \neq 0$. The resulting (Park) transformation angle $\phik = \int\omegak\dt$ aligns the $dq$-reference frame with the estimated rotor flux linkage $\psirdhat$. 
		\begin{remark}
			Note that magnetic saturation (i.e. varying $\Lm$ and $\Lr$) and parameter uncertainties (e.g. the temperature dependent rotor resistance $\Rr$) lead to an angle mismatch between the actual and estimated rotor flux angle, which results in the coupling of the flux and torque producing current components. However, with the proposed estimation scheme it is possible to define unique operating points (combinations of $\isd$ and $\isq$) that are reproducible and, thus, may serve as base vectors for efficiency or flux linkage LUTs.
			\label{rmrk:angle_misalignment}
		\end{remark}

	\subsubsection{Current PI-controllers with anti-windup and feed-forward compensation (FFC in Fig.~\ref{fig:control_system_overview})}
	%
	The control law $\usdqref = \usdqrefPI + \usdqrefFF \in\R^2$ comprises two parts, namely (i) the PI-controller $\usdqrefPI \in\R^2$ and (ii) the feed-forward controller $\usdqrefFF\in\R^2$ (see e.g.~\cite{Hackl2018, isie2016}). The PI-controller is given by 
	\begin{equation}
		\left.\begin{split}
		\usdqrefPI &= \Kpis \eobsdq[\isplain] + \Kiis \mv{\xi}_\symcurrent^{\csgdq}, \\
		\ddt\mv{\xi}_\symcurrent^{\csgdq} &= f_{\hat{u}}(\|\usdqref\|)\eobsdq[\isplain],
		\end{split}\quad\right\}
	\end{equation} 
	with proportional and integral gains $\Kpis\in\R$ and $\Kiis\in\R$ (identical for both current components), tracking error $\eobsdq[\isplain] := \isdqref - \isdq\in\R^2$, integrator output $\mv{\xi}_\symcurrent^{\csgdq} \in\R^2$  and anti-windup decision function	$f_{\hat{u}}(\cdot)$ as defined in \eqref{eq:anti_windup_decision_function}.
	The feed-forward part cancels out the back electromotive force (BEMF) terms, thus improving the dynamic response of the system. The BEMF terms can be compensated by chosing
	\begin{equation}
		\usdqrefFF = \omegak\sigma\Ls\J\isdq  + \tfrac{\Lm}{\Lr} (\omegar\J - \tfrac{1}{\Tr}\I2)\psirdqhat.
		\label{eq:feed-forward_control_equation}
	\end{equation}
	
	Naturally, the output voltage of the inverter is limited by the employed modulation strategy and the DC-link voltage $\udc$, e.g. using space-vector modulation (SVM) the voltage limit is given by $\hat{u}(\udc) = \tfrac{\udc}{\sqrt{3}}$. As to prevent exceeding this limit the voltage command is constrained by 
	\begin{equation}
		\usdqsatref = \left\{\begin{array}{ll}\usdqref,&  \|\usdqref\| \leq \hat{u}(\udc), \\  \usdqref\cdot\tfrac{\hat{u}}{\|\usdqref\|},& \text{else.}\end{array}\right.
	\end{equation}
	before it is passed on to the modulator. Finally, using conditional integration, a simple anti-windup decision function can be defined as
	\begin{equation}
		\resizebox{.89\linewidth}{!}{$%
		 f_{\hat{u}}: \R_{\geq0} \rightarrow \{0,1\}, \quad  f_{\hat{u}} := \left\{\begin{array}{ll}  1,&  \|\usdqref\| \leq \hat{u}(\udc), \\ 0,&  \text{else.}\end{array}\right.$}
		 \label{eq:anti_windup_decision_function}
	\end{equation}
		
	\subsection{Torque (feed-forward) control}
	
	The torque controller maps the torque reference $\melref$ to the stator current references $\isdqref$ by means of LUTs (see Fig.~\ref{fig:control_system_overview}) obtained from the machine identification process described in Sec.~\ref{sec:machine_identification}. The different torque control strategies are further elaborated in Sec.~\ref{sec:efficiency_analysis}.

\section{Machine identification}
\label{sec:machine_identification}

	The extraction of the exact MEPT curves requires machine identification, which goes beyond the standard \emph{no-load} and \emph{locked-rotor} tests. Assuming basic parameter knowledge and the control system implemented as described in the previous section, the advanced identification can be conducted as described in the following. The chart in Fig.~\ref{fig:identifaction_process_chart} shows the machine identification process.

	\begin{figure}
		\begin{tikzpicture}[every node/.style={minimum width=3cm, minimum height=1cm,inner sep=2mm, align=center, draw, rectangle, font=\footnotesize}, node distance=1cm and 1cm,scale=.8,transform shape]

			\node (basic_tests) []	{\textbf{Basic identification}\\ Perform locked-rotor \\ \& no-load tests\\ (see e.g.~\cite{ieee2018})}; 
			\node (ctrl_system) [right=of basic_tests]	{\textbf{Controller setup}\\ Implement control \\ system on DSP \\ (see Sec.~\ref{sec:control_system})}; 
			\node (measurement) [right=of ctrl_system]	{\textbf{Measurement}\\ Sweep through grid \\  and record data \\ (see Sec.~\ref{sec:measurement})}; 
			\node (map_extraction) [below=of measurement]	{\textbf{Data extraction} \\ Create machine maps \\ from processed data\\ (see Sec.~\ref{sec:map_extraction})};
			\node (lut_extraction) [left=of map_extraction]	{\textbf{LUT generation} \\  Generate LUTs from\\ extracted data maps\\ (see~Sec.~\ref{sec:lut_generation})}; 
			\node (mept_operation) [left=of lut_extraction, jkc3,dashed]	{\textbf{MEPT operation} \\  Run machine at\\ maximum efficiency\\ using MEPT LUTs}; 
			
			\draw[->] (basic_tests) -- (ctrl_system);
			\draw[->] (ctrl_system) -- (measurement);
			\draw[->] (measurement) -- (map_extraction);
			\draw[->] (map_extraction) -- (lut_extraction);
			\draw[->] (lut_extraction) -- (mept_operation);
		\end{tikzpicture}
		\caption{Generic machine identification process chart.}
		\label{fig:identifaction_process_chart}
	\end{figure}	
		
	\subsection{Setup description}
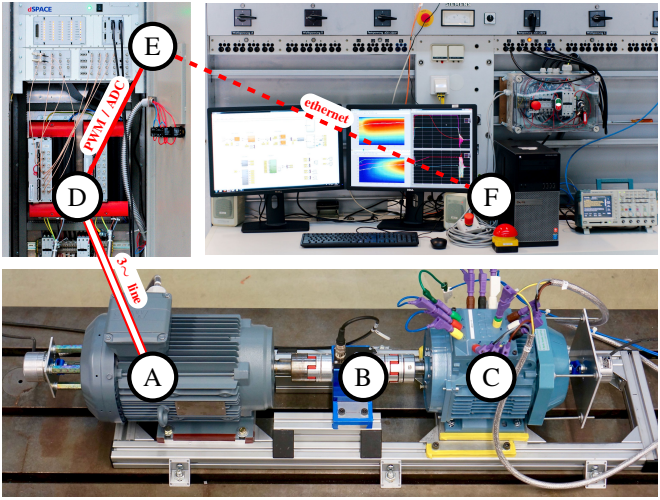
\begin{figure}
		\centering
		
 				\begin{tikzpicture}
 					\node[anchor=south west,inner sep=0] (image) at (0,-1) {\includegraphics[width=8.7cm , trim={0 0 0 7cm}, clip]{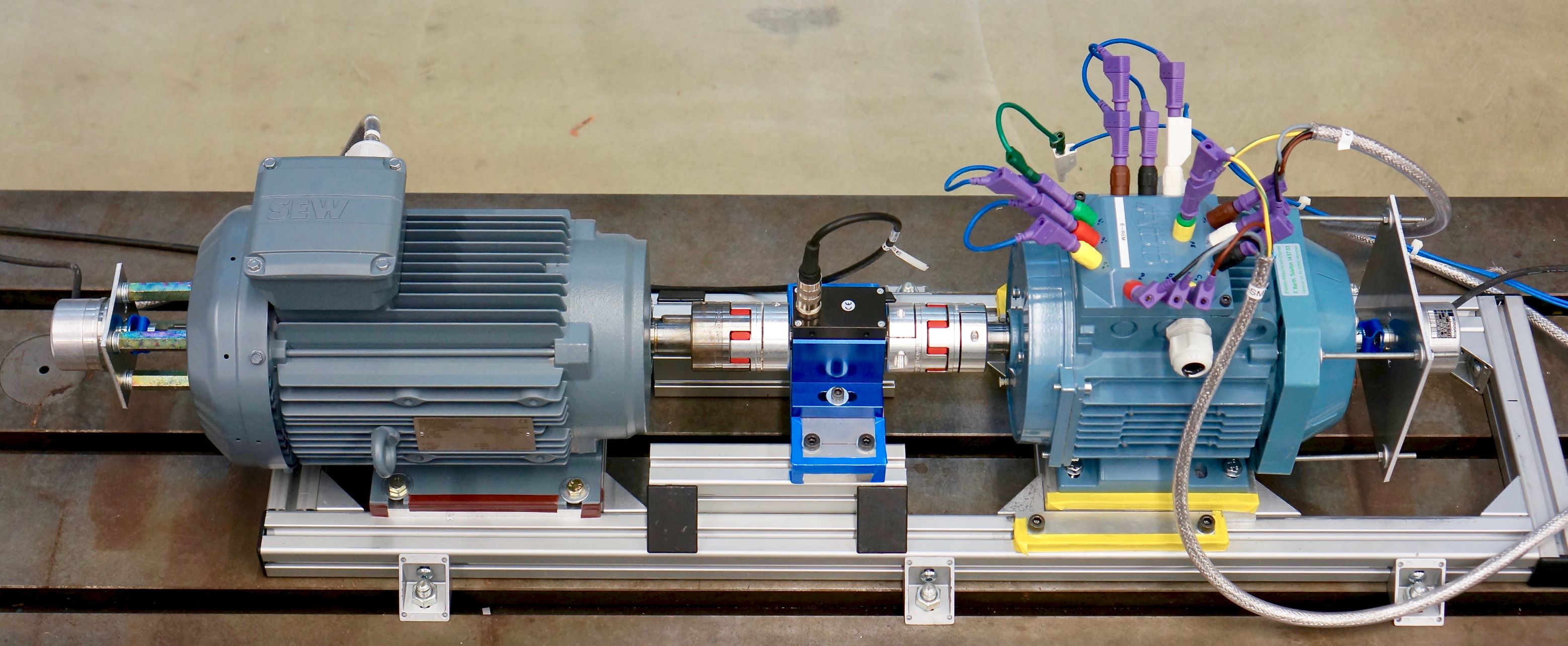}};
 					
 					\node[anchor=south west,inner sep=0] (image) at (0,2.2) {\includegraphics[width=2.5cm]{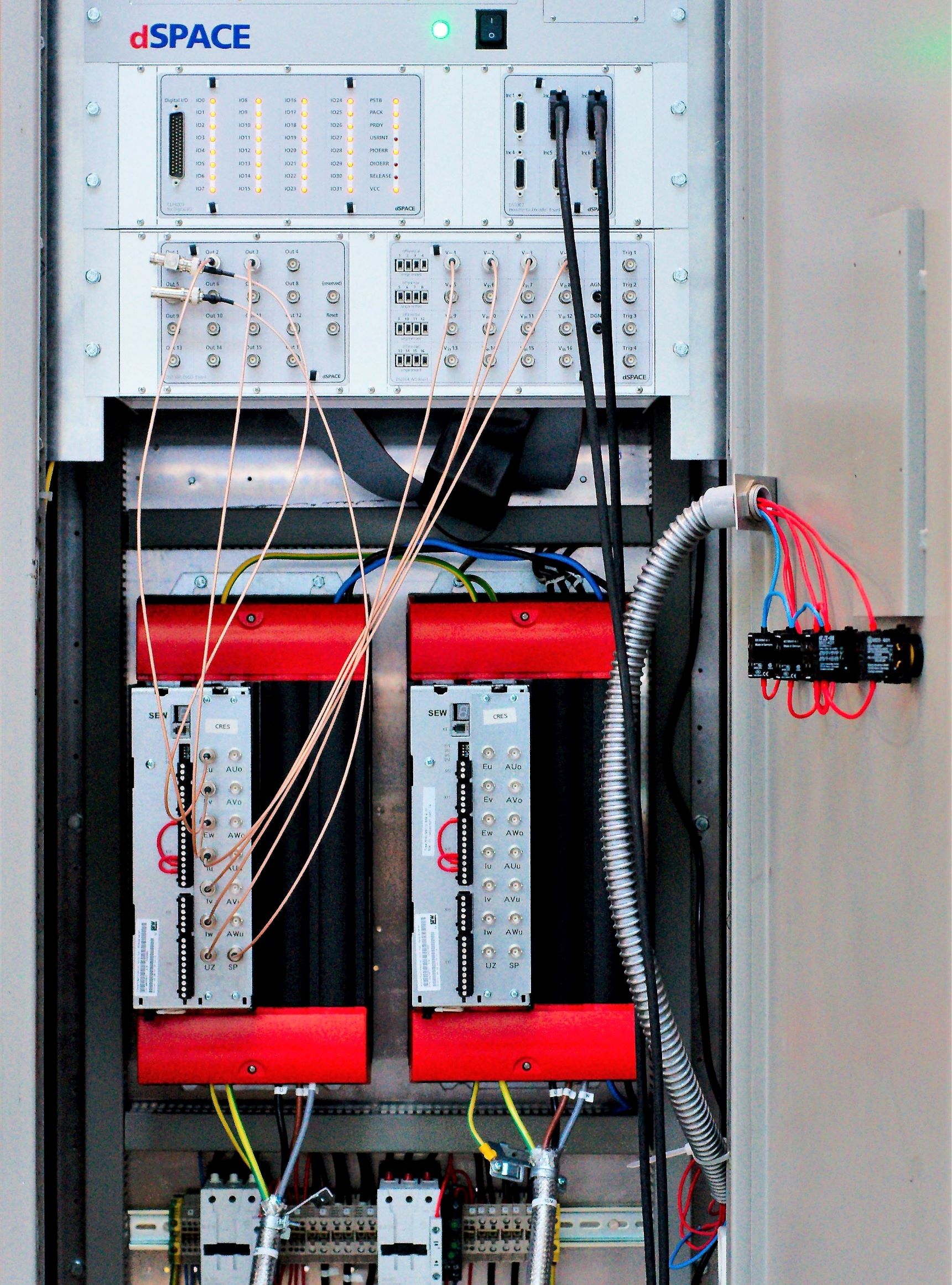}};
 					
 					\node[anchor=south west,inner sep=0] (image) at (2.7,2.225) {\includegraphics[width=6cm]{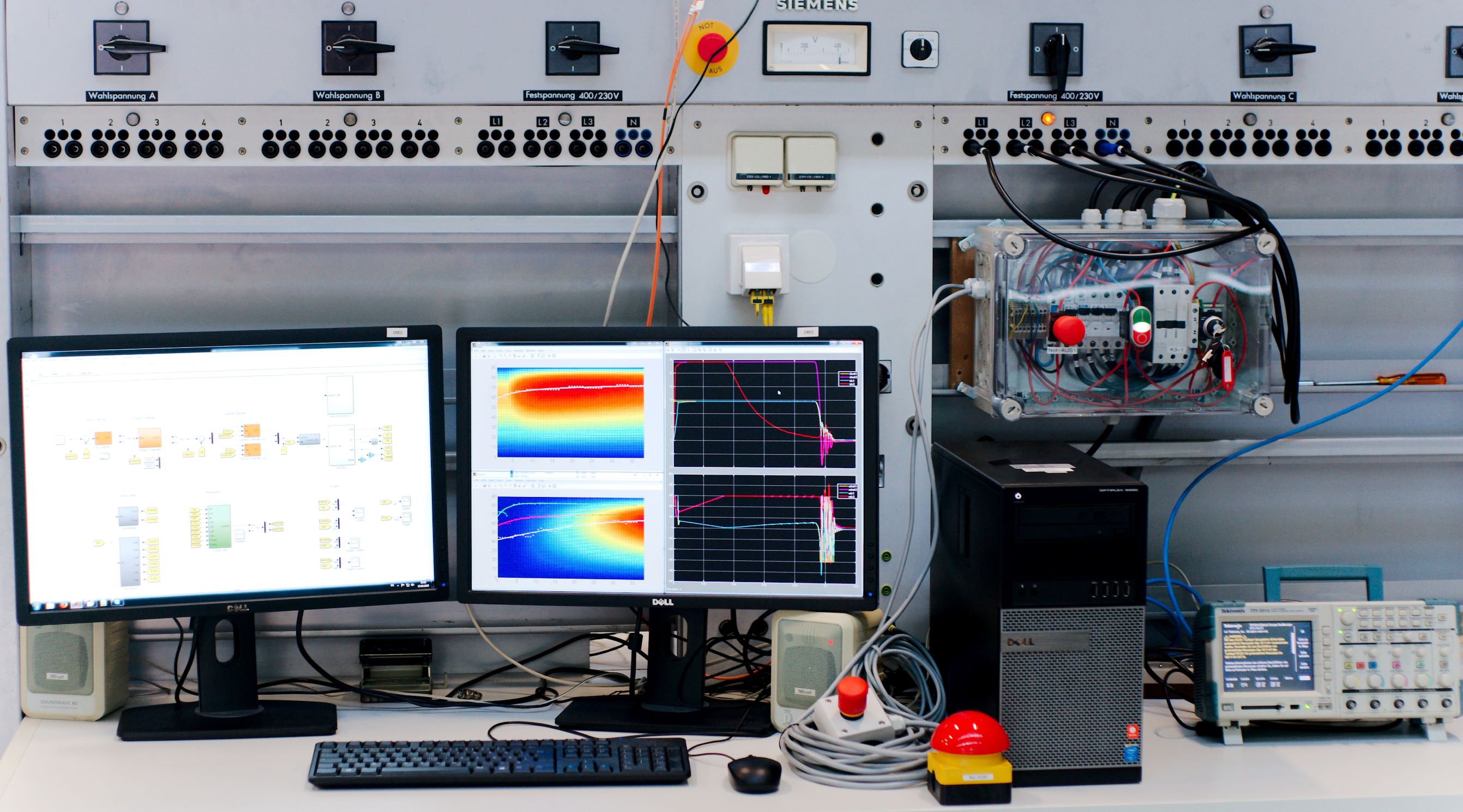}};
 					
 					\node[draw, circle, inner sep=1mm, fill=white, ultra thick] (A) at (2,0.6) {A};
 					\node[draw, circle, inner sep=1mm, fill=white, ultra thick] at (4.8,0.6) {B};
 					\node[draw, circle, inner sep=1mm, fill=white, ultra thick] at (6.5,0.6) {C};
 					\node[draw, circle, inner sep=1mm, fill=white, ultra thick] (D) at (1,3) {D};
 					\node[draw, circle, inner sep=1mm, fill=white, ultra thick] (E) at (2,5) {E};
 					\node[draw, circle, inner sep=1mm, fill=white, ultra thick] (F) at (6.5,3) {F};
 					\draw[dline,red] (D) --  node[pos=.5,above, sloped, fill=white, inner sep=2pt,rounded corners,yshift=2pt] {\tiny{\textbf{3$\sim$\, line}}}  (A);
 					\draw[dashed, red, ultra thick] (E) -- node[pos=.5,above, sloped, fill=white, inner sep=2pt,rounded corners] {\tiny \textbf{ethernet}} (F);
 					\draw[solid, red, ultra thick] (D) -- node[pos=.5,above, sloped, fill=white, inner sep=2pt,rounded corners] {\tiny{\textbf{PWM / ADC}}} (E);
 				\end{tikzpicture}
 				\caption{Overview of the test setup comprising (A) induction machine, (B) torque sensor, (C) load machine, (D) inverter, (E) real-time system and (F) host PC.}
		\label{fig:overview_drive_system}
	\end{figure}

	The components of the measurement setup are depicted in Fig.~\ref{fig:overview_drive_system}: A two-level voltage source inverter (VSI) with (measured) DC voltage input $\udc>0$ is connected to a star-connected three-phase SCIM. The stator voltages $\usabc\in\R^3$ are obtained from the inverter reference voltages and by assuming an inverter delay of one sampling period (regular sampled, symmetric PWM), whereas the stator currents $\isabc\in\R^3$ are measured directly.
	 The control system is implemented on a dSPACE real-time system, running at a fixed sampling frequency of \SI{4}{\kilo\hertz} and providing the space-vector modulated gate signals to the inverter. The machine is rotating at angular velocity $\omegam\neq0$ (measured), while the electrical frequency $\omegak$ is load dependent. The machine torque $\mel$ (measured), acts against the load torque $\mld\in\R$ and the frictional torque $\mfric\in\R$. Moreover, the load torque is produced by the load machine, which is speed controlled, thus keeping the angular velocity $\omegam$ at a constant reference value and allowing for torque variations of the IM. The respective system parameters (as given in the datasheet) are listed in Table~\ref{tab:parameters}.
	
		\begin{table}
			\caption{Parameters of the test setup}
			\centering
			\label{tab:parameters}
			\begin{tabular}{rlcrl}
				\toprule
				\parbox[r]{.1\textwidth}{}
					& \parbox[l]{.35\columnwidth}{\textbf{Parameter}}	
					& \parbox[l]{.1\columnwidth}{\textbf{Variable}}
					& \parbox[r]{.2\columnwidth}{\raggedleft\textbf{Value}}
					& \parbox[l]{.05\columnwidth}{\textbf{Unit}}\\
					
				\midrule
				
				\multirow{2}{*}{\rotatebox[origin=c]{90}{\textbf{VSI}}}	
							& DC-link voltage		& $\udc$	& 580 		& \si{\volt}			\\
							& Switching frequency	& $\fs$	& 4000		& \si{\hertz}		\\
		
				\midrule		
    				\multirow{12}{*}{\rotatebox[origin=c]{90}{\textbf{Induction machine}}}
    							& Rated speed 	& $\symangvel_{\idxmech,\idxrated}$	& 298.4			& \si{\radian\per\second}		\\	
    							& Rated torque 	& $\melnom$		& 10.05			& \si{\newton\meter}		\\				
    							& Rated voltage (amplitude)	& $\hat{\symvolt}_{\idxstator,\idxrated}$		& 327			& \si{\volt}		\\
    							& Rated current (amplitude)	& $\hat{\symcurrent}_{\idxstator,\idxrated}$	& 8.1		& \si{\ampere}		\\
    							& Rated flux (amplitude)		& $\hat{\symfluxl}_{\idxrotor,\idxrated}$	& 1.2 	& \si{\weber}		\\
    							& Number of pole pairs	& $\np$	& 1				& 		\\
    							& Stator resistance		& $\Rs$	& 2.3			& \si{\ohm}			\\
    							& Rotor resistance		& $\Rr$	& 1.55			& \si{\ohm}			\\
    							& Main inductance		& $\Lm$	& \num{340e-3}		& \si{\henry}   		\\
    							& Stator leakage inductance	& $\Lssi$				& \num{16.5e-3}		& \si{\henry}			\\
    							& Rotor leakage inductance 	& $\Lrsi$					& \num{16.5e-3}		& \si{\henry}	\\
    							& Moment of inertia  	& $\Thetam$					& \num{9.56e-3}		& \si{\kilo\gram\meter\squared}				\\
    							
    				\midrule
    		
    				\multirow{2}{*}{\rotatebox[origin=c]{90}{\textbf{PI}}}				
    							& P-gain 	& $\Kpis$		& 0.8			& \si{\ohm} \\
    							& I-gain    & $\Kiis$	& 136	& \si{\ohm\per\second} \\ 
    				\bottomrule
    			\end{tabular}
    		\end{table}
    
	\subsection{Measurement methodology}
	\label{sec:measurement}
	Since the load machine is speed controlled, the IM currents can be set arbitrarily without accelerating the rotor. The fundamental idea is that each combination of stator currents $(\isd, \isq)$ represents a reproducable and unique operating point, and, hence, the objective is to gather machine information for each pair of currents within the feasible range, i.e. $(\isd, \isq) \in \mathbb{I_\idxstator} = \{ (\isd,\isq)\, | \, {\isd}^2 + {\isq}^2 \leq \isnom^2 \}$. This is achieved by sampling the current plane on a regular grid and sweeping through each grid point, whilst recording measured and estimated data, such as stator voltages, currents, rotor flux linkages, speeds and angles. In order to avoid redundancy in the measurements, only the positive $d$-axis is covered. Moreover, while a minimum amount of excitation current $\isdmin \in \R$ is required for the rotor flux orientation to operate properly, magnetic saturation sets in for high values of $\isd$, which motivates for an upper limit of half the rated current, i.e. $\isd \in [\isdmin, \tfrac{1}{2}\isnom]$. In turn, the $q$-current is varied from negative rated to positive rated current, i.e. $\isq \in [-\isnom,\isnom]$. Now, the grid vectors can be defined over the given intervals as 
	\begin{equation*}
		\begin{split}
			\mv{\isdgrid} &= (\isdgrid[1]=\isdmin, \ldots, \isdgrid[j], \ldots, \isdgrid[m] = \tfrac{1}{2}\isnom),\\
			\mv{\isqgrid} &= (\isqgrid[1]=-\isnom, \ldots, \isqgrid[k], \ldots, \isqgrid[n]=\isnom), 
		\end{split}
	\end{equation*}	
	for $1\leq j \leq m$ and $1\leq k \leq n.$. The numbers of sampling points $m,n\in \N$ per current component define the grid size ($m\times n$) and should be a trade-off between accuracy and measurement effort. Note that the grid corners are located outside of $\mathbb{I}_\idxstator$ and, hence, the stator currents exceed the rated value.

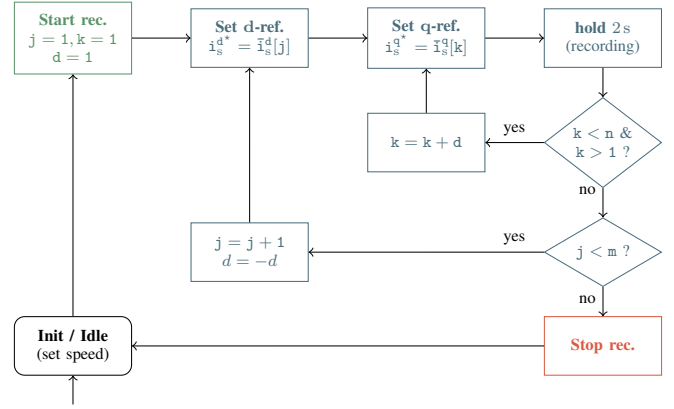
\begin{figure}
		\resizebox{\linewidth}{!}{
		\begin{tikzpicture}[nnode/.style={minimum width=2cm, minimum height=1cm,inner sep=2mm, align=center, draw, rectangle, font=\footnotesize}, node distance=.5cm and 1cm,scale=.8,transform shape]

			\node (start) 			[nnode,jkc3]			 													{\textbf{Start rec.}\\ $\mathtt{j=1, k=1}$\\ $\mathtt{d=1}$};
			\node (set_isd) 		[nnode,right=of start,jkc1]		 											{\textbf{Set $\mathbf{d}$-ref.} \\ $\mathtt{\isdref = \isdgrid[j]}$};	
			\node (set_isq) 		[nnode,right=of set_isd,jkc1]	 											{\textbf{Set $\mathbf{q}$-ref.} \\ $\mathtt{\isqref = \isqgrid[k]}$};			
			\node (record) 		[nnode,right=of set_isq,jkc1]	 											{\textbf{hold} \SI{2}{\second} \\ (recording)};		
			\node (j_if) 			[nnode,below=of record, diamond, inner sep=0mm,jkc1]	 	{$\mathtt{k < n} $ \&\\ $\mathtt{k>1}$ ?};			
			\node (i_if) 			[nnode,below=of j_if, diamond, inner sep=0mm,jkc1]	 		{$\mathtt{j < m}$ ?};				
			\node (j_iter) 		[nnode,left=of j_if,jkc1]	 													{$\mathtt{k=k+d}$};			
			\node (i_iter) 		[nnode,jkc1] at (set_isd |- i_if) 												{$\mathtt{j=j+1}$\\ $d=-d$};		
			\node (stop) 			[nnode,below=of i_if, jkc2]											{\textbf{Stop rec.}};		
			\node (idle) [nnode] at (start |- stop)	[rounded corners]					{\textbf{Init / Idle} \\ (set speed)};
			
			\draw [->]		(start) 		-- 		(set_isd);
			\draw [->]		(set_isd) 	-- 		(set_isq);
			\draw [->]		(set_isq) 	-- 		(record);
			\draw [->]		(record) 	-- 		(j_if);
			\draw [->]		(j_if) node[above left,xshift=-1.2cm] {\footnotesize yes} 			-- 		(j_iter);
			\draw [->]		(j_iter) 		-- 		(set_isq);
			\draw [->]		(j_if) 	node[above left,yshift=-1cm] {\footnotesize no} 			-- 		(i_if);
			\draw [->]		(i_if) 	node[above left,xshift=-1.2cm] {\footnotesize yes} 			-- 		(i_iter);
			\draw [->]		(i_iter) 		-- 		(set_isd);
			\draw [->]		(i_if) 	 	node[above left,yshift=-1cm] {\footnotesize no} 		-- 		(stop);
			\draw [->]		(stop) 		-- 		(idle);
			\draw [->]		(idle) 		-- 		(start);
			\draw [->]		(idle) ++ (0,-1) 	-- 		(idle);
			
		\end{tikzpicture}
		} 
		\caption{Flow chart of the measurement procedure.}
		\label{fig:measurement_flow_chart}
	\end{figure}

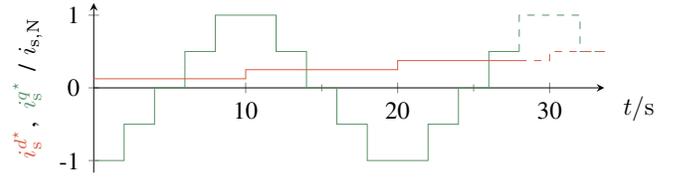
\begin{figure}
		\resizebox{\linewidth}{!}{%
		\begin{tikzpicture}
			\begin{axis}[
					grid style={line width=1pt, draw=gray},
					width=1\linewidth,
					height=4cm,
					axis lines=left, 
					xlabel=$t / \si{\second}$,
					xmin=0,xmax=8.4,ymin=-3.5,ymax=3.5,
					minor x tick num={1},
					axis x line=center,
					every axis x label/.style={
						at={(ticklabel* cs:1.02)},
						anchor=north west
					},
					ylabel={{\color{jkc2}$\isdref$},\, {\color{jkc3}$\isqref$} / $\isnom$},
					enlargelimits=false,
					xtick = {2.5, 5, 7.5},
					ytick = {-3, 0, 3},
					yticklabels = {-1, 0, 1},
					xticklabels = {10, 20, 30}
					]
				\addplot[domain=0:5,jkc3] coordinates {
						(0,-3) (0.5,-3) (0.5,-1.5) (1,-1.5) (1,0) (1.5,0) (1.5,1.5) (2,1.5) (2,3) (2.5,3) (3,3)	(3,1.5) (3.5,1.5) (3.5,0)	(4,0) (4,-1.5) (4.5,-1.5) (4.5,-3) (5.5,-3) (5.5,-1.5)
						(6,-1.5) (6,0) (6.5,0) (6.5, 1.5) (7,1.5)
						};
						
				\addplot[domain=0:5,jkc3,dashed] coordinates {
						(7,1.5)  (7,3) (8,3) (8,1.5) (8.5,1.5) (8.5,0)
						};
						
				\addplot[domain=0:5,jkc2] coordinates {
						(0,0.375) (2.5,0.375) (2.5,0.75) (5,0.75) (5,1.125) (7,1.125)
						};
					
				\addplot[domain=0:5,jkc2,dashed] coordinates {
						(7,1.125) (7.5, 1.125) (7.5, 1.5) (8.5,1.5)
						};
			\end{axis}
		\end{tikzpicture}
		}
		\caption{Illustration of the current reference sweeping.}
		\label{fig:current_sweeing}
	\end{figure}

	The measurement procedure is shown in Fig.~\ref{fig:measurement_flow_chart}: Initially, the speed command for the load machine is set to a constant value. Once the recording is started, the current grid is swept through with an idle time of e.g. \SI{2}{\second} at each operating point, allowing for the system to reach steady-state. Upon changes in $\isdref$, the direction of $\isqref$ is reversed, which avoids large torque jumps. When the whole grid has been covered, the recording is stopped. The current reference sweeping is further illustrated in Fig.~\ref{fig:current_sweeing}. Since the machine speed influences the efficiency of the machine, the measurements are repeated for different speeds, which constitutes an additional dimension in the LUTs.

\subsection{Extraction of $dq$-machine maps}
\label{sec:map_extraction}
	
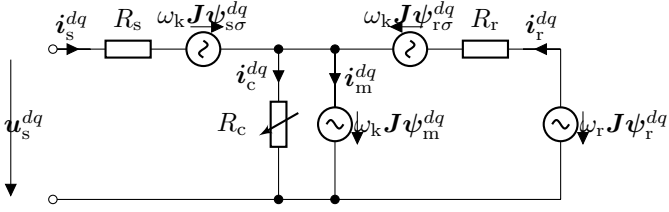
\begin{figure}
		\begin{circuitikz}[/tikz/circuitikz/bipoles/length=.75cm]
			\draw (0,0) coordinate (0) to[short, o-, i=$\isdq$] ++ (.5,0) 
							 to[R=$\Rs$] ++ (1,0) to[sV=$\omegak\J\psissidq$] ++ (1,0) 
							 to[short] ++ (0.5,0) coordinate (a) 
							 to[short] ++ (.75,0) coordinate (b) 
							 to[short] ++ (0.5,0) 
							 to[sV<=$\omegak\J\psirsidq$] ++ (1,0) 
							 to[R=$\Rr$] ++ (1,0) 
							 to[short, i<=$\irdq$] ++ (0.5,0) 
							 to[sV=$\omegar\J\psirdq$] ++ (0, -2) coordinate (c) 
							 to[short] (c -| b) coordinate (d) 
							 to[short] (c-|a) coordinate (e) 
							 to[short,-o] (c-|0) coordinate (f);

			\draw (a) to[vR, l_=$\Rfe$, *-*, i>_=$\ifedq$] (e);
			\draw (b) to[sV=$\omegak\J\psimdq$, *-*, i>^=$\imdq$] (d);
			\draw (0) to[open, v=$\usdq$] (f);
			
		\end{circuitikz}
		\caption{Steady-state T-equivalent circuit of the squirrel-cage induction machine (see e.g.~\cite{levi1996}).}
		\label{fig:equivalente_circuit_im_ss}
	\end{figure}

	The extraction of the machine characteristics described in the following is based on a steady-state evaluation of the recorded measurement data. In steady-state, the time-derivatives in \eqref{eq:machine_model} become zero and the equivalent circuit shown in Fig.~\ref{fig:equivalente_circuit_im_ss} holds. Prior to the data extraction, the measurement data has been postprocessed by (i) digital lowpass filtering with filter time constant $T_f = \SI{25}{\milli\second}$, (ii) cropping off the transient part of each sampling window and (iii) calculating the mean value of the remaining data points for each window. Moreover, since the measurement was conducted for two quadrants only, symmetry properties (see e.g.~\cite{isie2016}) have been exploited as to expand the data over all four quadrants. Finally, the 3D data maps have been smoothed using the \texttt{loess} curve fitting method in MATLAB R2018a.	
		
	\subsubsection{Flux linkage maps}
	%
	
	The stator flux linkage $\psisdq = \psissidq + \psimdq$, which is the sum of the stator leakage flux $\psissidq\in\R^2$ and air gap flux $\psimdq\in\R^2$, is obtained from the steady-state stator voltage equation in \eqref{eq:machine_model} by solving for $\psisdq$, i.e.
	\begin{equation}
		\psisdq \overset{\eqref{eq:machine_model}}{=} \tfrac{1}{\omegak}\J^{-1}\left( \usdq - \Rs\isdq\right).
		\label{eq:steady_state_stator_flux}
	\end{equation}
	
\begin{figure*}[!h]
		\centering
		\subfloat[]{
			\tikz{
				\node[inner sep=0] at (0,0) {\includegraphics[width=5.5cm]{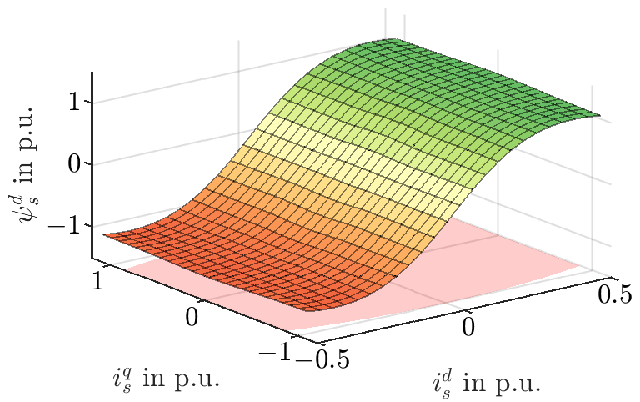}};
				\node[red!60] at (1.4,-0.5) {$\mathbb{I_\idxstator}$};
			}
			\label{fig:flux_linkage_d}
		}\quad
		\subfloat[]{
			\tikz{
				\node[inner sep=0] at (0,0) {\includegraphics[width=5.5cm]{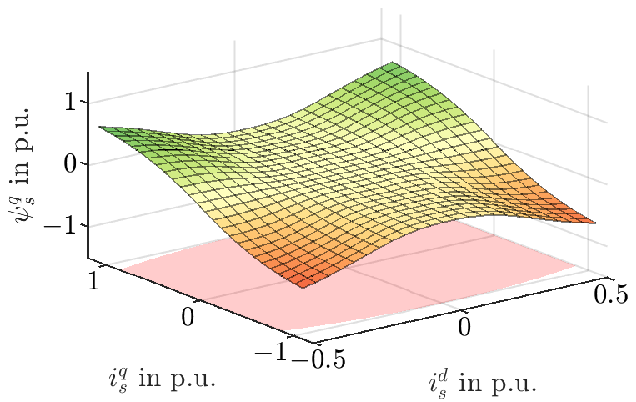}};
				\node[red!60] at (1.4,-0.5) {$\mathbb{I_\idxstator}$};
			}
			\label{fig:flux_linkage_q}
			}\quad
		\subfloat[]{
			\tikz{
				\node[inner sep=0] at (0,0) {\includegraphics[width=5.5cm]{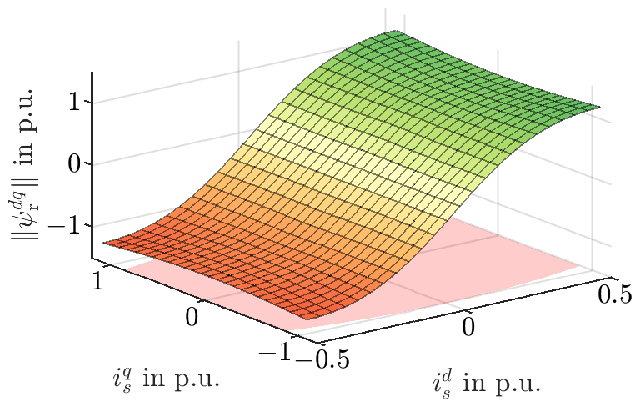}};
				\node[red!60] at (1.4,-0.5) {$\mathbb{I_\idxstator}$};
			}
			\label{fig:rotor_flux_linkage}
		}
		\caption{Flux linkage maps in the estimated rotor flux oriented $dq$-reference frame for constant speed $\omegam = \SI{150}{\radian\per\second}$: (a) stator $d$-component, (b) stator $q$-component and (c) rotor $d$-component.}	
		\label{fig:flux_linkages_3d}
	\end{figure*}

	
	\begin{remark}
		It should be noted that the stator resistance is temperature dependent. 
	 If the machine is heated up before the identification process, an approximate resistance value can be obtained from the measurement data: For zero torque, the stator flux linkage is aligned with the rotor flux linkage, hence giving $\psisq = \SI{0}{\weber}$ and $\Rs \approx \left.\tfrac{\usd}{\isd}\right|_{\isq=0}$.
		\label{rmrk:stator_resistance_temperature}
	\end{remark}
	The resulting flux maps are shown in Fig.~\ref{fig:flux_linkages_3d}. It can be seen in the $d$-stator flux linkage component (see Fig.~\ref{fig:flux_linkage_d}), that already for low values of $\isd$ (around $\isd = \SI{0.3}{p.u.}$), the flux linkage saturates almost completely. Moreover, it can be observed in the $q$-component (see Fig.~\ref{fig:flux_linkage_q}), that in close vicinity of zero excitation (around $\isd = \SI{0}{p.u.}$), $\psisq$ is almost zero, because the rotor flux linkage is in line with the stator flux linkage. Only for higher loads, there is an angle difference between the rotor and stator flux linkages, which causes a significant $q$-component of the stator flux linkage.

	 Likewise, the rotor flux linkage $\psirdq = \psirsidq + \psimdq$, with rotor leakage flux linkage $\psirsidq\in \R^2$, is calculated from the steady-state rotor voltage equation in \eqref{eq:machine_model} as
	\begin{equation}
		\psirdq \overset{\eqref{eq:machine_model}}{=} -\tfrac{\Rr}{(\omegak-\omegar)}\J^{-1}\irdq.
		\label{eq:rotor_flux_linkage_steady_state}
	\end{equation}
	Since the rotor currents cannot be measured directly in a SCIM, the determination of $\psirdq$ is not straightforward. 
	%
	However, multiplying both sides from the left with ${\psirdq}^\top$ yields 
	\begin{equation}
		\|\psirdq\|^2 = -\tfrac{\Rr}{\omegak-\omegar}{\psirdq}^\top\J^{-1}\irdq,
	\end{equation}
	which is proportional to the machine torque [see \eqref{eq:machine_torque}, $\dagger$ term]. Taking the square root and inserting \eqref{eq:machine_torque} finally gives
	%
	%
	\begin{equation}
		\|\psirdq\| 
		 \overset{\eqref{eq:machine_torque}}{=}  \sqrt{\tfrac{\Rr}{\omegak-\omegar}\tfrac{2}{3\np}\mel}.
		\label{eq:rotor_flux_linkage_2}
	\end{equation}
	
	Owing to \eqref{eq:rotor_flux_linkage_ideal}, the $q$-component of the rotor flux is zero.

	\begin{remark}
		Note that \eqref{eq:rotor_flux_linkage_2} depends on the temperature dependent rotor resistance $\Rr$ and should be adapted together with $\Rs$ (see~Remark~\ref{rmrk:stator_resistance_temperature}). Moreover, \eqref{eq:rotor_flux_linkage_2} is not defined for zero torque (i.e. $\omegak-\omegar = 0$) and, hence, interpolation is required in the respective range. Since the square root yields positive values only, it is furthermore necessary to exploit the symmetry properties of the rotor flux linkage (see e.g.~\cite{isie2016}).
	\end{remark}
	
	The resulting rotor flux map is shown in Fig.~\ref{fig:rotor_flux_linkage}, showing the magnitude of the rotor flux linkage $\psirdq$. Similar to the stator $d$-component, the effect of magnetic saturation can be clearly observed in $d$-direction. Moreover, a slight variation is observed in $q$-direction, which is due to magnetic coupling between the two axes.  
	
	\subsubsection{Machine torque map}
\begin{figure}[!h]
		\centering
		\tikz{
			\node[inner sep=0] at (0,0) {	\includegraphics[width=8.5cm]{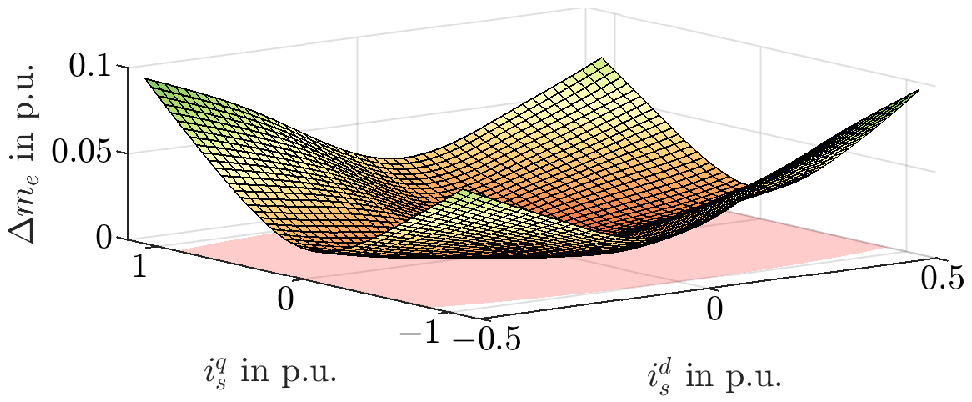}};
			\node[red!60] at (2.3,-0.35) {$\mathbb{I_\idxstator}$};
		}
		\caption{Torque difference $\Delta\mel$ between measured and reconstructed torque for constant speed $\omegam = \SI{150}{\radian\per\second}$.}	
		\label{fig:torque_lut_3d_4Q}
	\end{figure}

	Since the rotor current and flux linkage cannot be measured, the machine torque has to be either measured using a torque sensor, or approximated by \eqref{eq:machine_torque} [$\ddagger$ term], assuming $\mfe = 0$. The measured torque $\mme \in \R$ includes mechanical losses (e.g. friction), though, which can be eliminated: The speed-dependent friction torque $\mfric$ is equal to the measured torque for reference currents $(\isdmin, 0)$, as the machine torque should be very close to zero at this point. Hence, calculating $\mel = \mme - \mfric$ gives the machine torque. Alternatively, the machine torque can be approximated using the measured stator currents and calculated stator flux linkage, i.e. $\melhat = \tfrac{3}{2}\np{\isdq}^\top\J\psirdq$. Apart from neglecting iron losses, the downside of the reconstruction is its dependency on the stator resistance and the stator voltage estimate, which are required for the flux linkage calculation (see Remark~\ref{rmrk:stator_resistance_temperature}). The torque difference $\Delta\mel = \melhat-\mel$ is shown in Fig.~\ref{fig:torque_lut_3d_4Q}. Moreover, the first quadrant of the measured torque map, together with the trajectories of the different torque control strategies (more details given in Sec.~\ref{sec:lut_generation}) is shown in Fig.~\ref{fig:torque_lut_3d}.
	
	\subsubsection{Efficiency map}
	\label{sec:power_balance}
	The steady-state T-equivalent circuit of the SCIM considering core losses is depicted in Fig.~\ref{fig:equivalente_circuit_im_ss}. The (variable) core loss resistance $\Rfe>0$ is driven by the current $\ifedq\in\R^2$. Moreover, the magnetizing current $\imdq\in\R^2$ runs through the branch parallel to the core resistance. 
	In the following (using amplitude correct Clarke transformation) let
	\begin{equation}
		\resizebox{.89\linewidth}{!}{$%
			\pel = \tfrac{3}{2}{\isdq}^\top\usdq, \quad
			\pme = \mel\omegam, \quad
			\ploss = p_{\mathrm{cu,s}} + p_{\mathrm{cu,r}} + p_{\mathrm{fe}}
		$}
		\label{eq:power_definitions}
	\end{equation}
	be the electrical power ($\pel$) at the machine terminals, the mechanical power ($\pme$) transmitted via the shaft and the sum of losses ($\ploss$) with stator copper losses $p_{\mathrm{cu,s}}>0$,  rotor copper losses $p_{\mathrm{cu,r}}$ and iron losses $p_{\mathrm{fe}}$, respectively. The power equilibrium requires that the (active) electrical power at the machine terminals is equal to the sum of all active power terms in the circuit, i.e.
	\begin{equation}
		\begin{split}
			\pel
				&= \ploss +  \tfrac{3}{2}\omegak{\isdq}^\top\J\psisdq	+  \tfrac{3}{2}\omegak{\irdq}^\top\J\psirdq\\
				&\quad - \tfrac{3}{2}\omegak{\ifedq}^\top\J\psimdq  + \pme.
		\end{split}
	\end{equation}
	Since $\pel=\ploss+\pme$, it can be concluded, that the following must hold for the residual power terms
	\begin{equation}
		-\tfrac{3}{2}\omegak{\irdq}^\top\J\psirdq = \tfrac{3}{2}\omegak{\isdq}^\top\J\psisdq - \tfrac{3}{2}\omegak{\ifedq}^\top\J\psimdq.
	\end{equation}
	Multiplication by $\np/\omegak$ gives the torque equivalence~\eqref{eq:machine_torque}, with 
	\begin{equation}
		\mfe =  \tfrac{3}{2}\np{\ifedq}^\top\J\psimdq.
	\end{equation}
	
	If $\pel\geq 0$ (per definition, passive sign convention), the machine operates in motor mode, while for $\pel<0$ it operates as a generator. Hence, the efficiency is defined as
	
	\begin{equation}
		\eta := \left\{\begin{array}{ll} \tfrac{\pme}{\pel}&,\quad \text{for}\quad \pel \geq 0 \\ \tfrac{\pel}{\pme}&, \quad \text{for} \quad \pel < 0. \end{array}\right.
	\end{equation}
	
	Note that the efficiency assesses the electromagnetic conversion process in the machine, whereas friction and mechanical losses are \emph{not} considered. The first quadrant of the efficiency map is shown in Fig.~\ref{fig:eta_lut_3d}.
	
	\begin{figure*}[!h]
		\centering
		
		\begin{tikzpicture}[every node/.style={font=\footnotesize\bfseries, inner sep=0}]
			\draw[very thick, draw=plca3, dashed] (0,0) -- (0.75,0) node[right, xshift=1mm, anchor=west] (vhz_rated) {V/Hz (rated)};
			\draw[very thick, draw=plca2, dash dot] (vhz_rated.east) ++ (0.5,0) --  ++ (0.75,0) node[right, xshift=1mm, anchor=west] (vhz_opot) {V/Hz (opt.)};		
			\draw[very thick, draw=plca1, dotted] (vhz_opot.east) ++ (0.5,0) --  ++ (0.75,0) node[right, xshift=1mm, anchor=west] (constant_flux) {Constant Flux};		
			\draw[very thick, draw=plca4] (constant_flux.east) ++ (0.5,0) --  ++ (0.75,0) node[right, xshift=1mm, anchor=west] (mtpa) {MTPC ($\mel$)};	
			\draw[very thick, draw=plca5] (mtpa.east) ++ (0.5,0) --  ++ (0.75,0) node[right, xshift=1mm,anchor=west] (mept) {MEPT ($\mel$)};	
			\draw[] (-0.2,-0.3) rectangle ($(mept.north east) + (0.3,0.2)$);
\end{tikzpicture}
\vspace{0cm}
		
		\subfloat[]{
			\tikz{
				\node[inner sep=0] at (0,0) {\includegraphics[width=5.5cm]{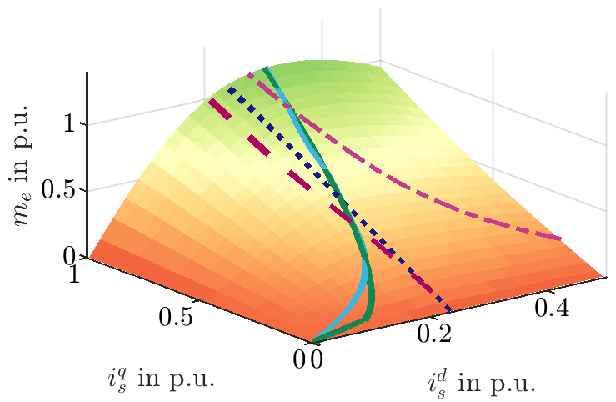}};
			}
			\label{fig:torque_lut_3d}
		}\quad
		\subfloat[]{
			\tikz{
				\node[inner sep=0] at (0,0) {\includegraphics[width=5.5cm]{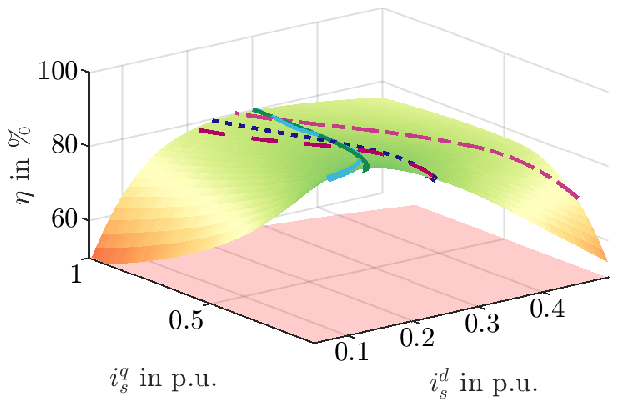}};
				\node[red!60] at (1.4,-0.55) {$\mathbb{I_\idxstator}$};
			}
			\label{fig:eta_lut_3d}
		}\quad
		\subfloat[]{
			\tikz{
				\node[inner sep=0] at (0,0) {\includegraphics[width=5.5cm]{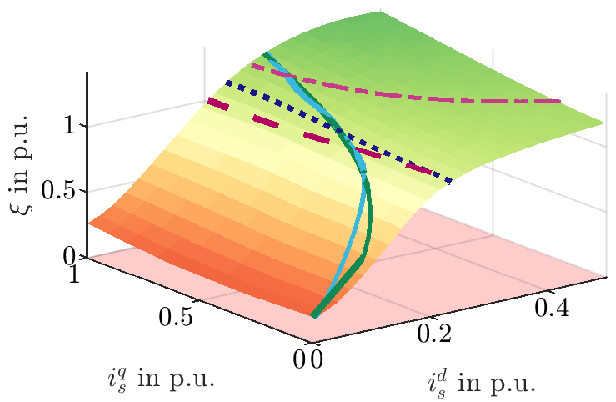}};
				\node[red!60] at (1.4,-0.55) {$\mathbb{I_\idxstator}$};
			}
			\label{fig:uf_lut_3d}
		}
		\caption{First quadrant machine maps in the rotor flux oriented $dq$-reference frame for constant speed $\omegam = \SI{150}{\radian\per\second}$: (a) torque, (b) efficiency and (c) voltage over frequency (V/Hz) ratio.}	
	\end{figure*}
	
	The electrical power losses of the machine are found in the resistive components, i.e. (i) the stator resistance $\Rs$, (ii) the rotor resistance $\Rr$ and (iii) the core resistance $\Rfe$. Note that the mentioned resistances are temperature dependent (see Remark~\ref{rmrk:stator_resistance_temperature}), and, thus, losses vary with the temperature as well. The ohmic losses in the stator are given by
	\begin{equation}
		p_{\mathrm{cu,s}} = \tfrac{3}{2}\Rs{\isdq}^\top\isdq.
	\end{equation}
	
	The rotor ohmic losses can be calculated by replacing $\psirdq$ in \eqref{eq:machine_torque} by \eqref{eq:rotor_flux_linkage_steady_state}, i.e. 
	\begin{equation}
		\mel  = \tfrac{3}{2}\np\tfrac{\Rr}{\omegak-\omegar}{\irdq}^\top\irdq,
	\end{equation}
	which can then be used to express the rotor losses as
	\begin{equation}
		p_{\mathrm{cu,r}} = \tfrac{3}{2}\Rr{\irdq}^\top\irdq = \tfrac{1}{\np}(\omegak-\omegar)\mel.
	\end{equation}
	
	\begin{remark}
		Note that neither the rotor currents $\irdq$, nor the rotor resistance $\Rr$ show up in the power loss term, which makes the calculation less error prone and, hence, more reliable and accurate, supposed that the stator resistance required for the flux calculation is determined correctly. 
	\end{remark}
	
	Lastly, the core losses are obtained by inserting the previous results into \eqref{eq:power_definitions} with $\pel = \pme + \ploss$, i.e.
	\begin{equation}
		p_{\mathrm{fe}} = \tfrac{3}{2}{\ifedq}^\top\ifedq = \pel-\pme- p_{\mathrm{cu,s}} -p_{\mathrm{cu,r}}.
	\end{equation}
	

	\subsubsection{V/Hz map}
	\label{sssec:vhz_map}
	The recorded data further allows for calculating a V/Hz ratio map, which would typically be difficult to measure for a constant speed, due to the effect of slip. Since constant speed is assured by the load machine here, the V/Hz ratio, in the following denoted by $\xi \in \R$, is calculated by 
	\begin{equation}
		\xi := \left(2\pi\sqrt{{\usd}^2 + {\usq}^2}\right)/\omegak.
	\end{equation}

		\subsection{Look-up table generation}
		\label{sec:lut_generation}
		Since the mapping of the machine torque $\mel$ to the stator currents $\isdq$ is ambiguous, different torque control strategies may be applied, producing equal torque output, while being subject to either an equality constraint or an optimization problem (see e.g.~\cite{eldeeb2017}). The general procedure for the LUT generation is to (i) calculate torque contour lines for reference torque values $\melref \in [-\melnom,\melnom]$, (ii) use the resulting $d$- and $q$-currents for looking up the secondary variable from the respective map (e.g. $\eta$ or $\xi$) and (iii) evaluate the equality constraint or the optimization problem on those values. The best match is selected and the corresponding $\isdref$ and $\isqref$ are stored in the LUTs $\mathcal{L}_{\isdref}^{\mathrm{M}}$ and $\mathcal{L}_{\isqref}^{\mathrm{M}}$ (with superscript 'M' being replaced by the respective torque strategy). Repeating this procedure for machine maps of different speeds, finally produces the 2D LUTs $\mathcal{L}_{\isdref}^{\mathrm{M}}(\melref,\omegam)$ and $\mathcal{L}_{\isqref}^{\mathrm{M}}(\melref,\omegam)$. In the following, the most common torque control strategies, as well as the proposed MEPT strategy are briefly introduced. The calculated LUTs are shown in Fig.~\ref{fig:LUTs}.
		
		\begin{figure*}[!h]
			\centering
			\subfloat[$\mathcal{L}_{\isdref}^{\mathrm{VHz}}(\melref,\omegam)$]{\includegraphics[width=4.5cm]{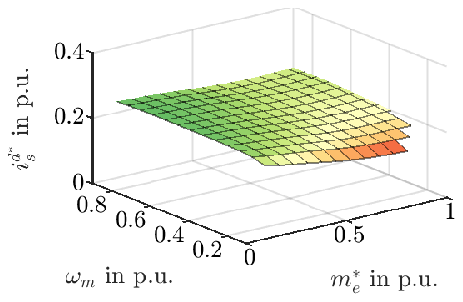}\label{fig:lisd_vhz}}
			\subfloat[$\mathcal{L}_{\isqref}^{\mathrm{VHz}}(\melref,\omegam)$]{\includegraphics[width=4.5cm]{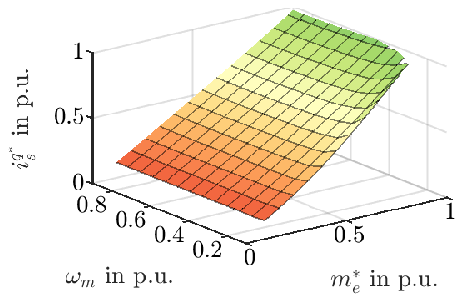}\label{fig:lisq_vhz}}
			\subfloat[$\mathcal{L}_{\isdref}^{\mathrm{CF}}(\melref,\omegam)$]{\includegraphics[width=4.5cm]{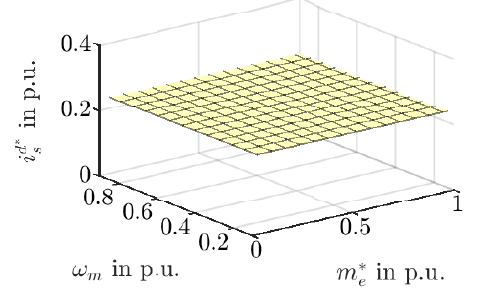}\label{fig:lisd_cf}}
			\subfloat[$\mathcal{L}_{\isqref}^{\mathrm{CF}}(\melref,\omegam)$]{\includegraphics[width=4.5cm]{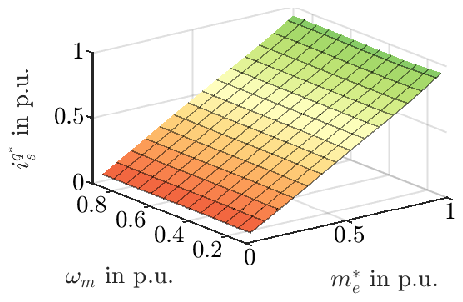}\label{fig:lisq_cf}}\\
			\subfloat[$\mathcal{L}_{\isdref}^{\mathrm{MTPC}}(\mel,\omegam)$]{\includegraphics[width=4.5cm]{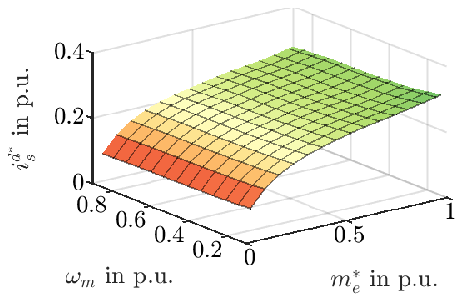}\label{fig:lisd_mtpc}}
			\subfloat[$\mathcal{L}_{\isqref}^{\mathrm{MTPC}}(\mel,\omegam)$]{\includegraphics[width=4.5cm]{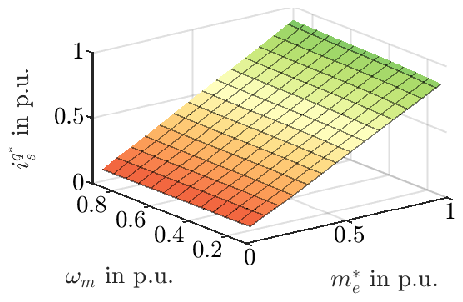}\label{fig:lisq_mtpc}}
			\subfloat[$\mathcal{L}_{\isdref}^{\mathrm{MEPT}}(\melref, \omegam)$]{\includegraphics[width=4.5cm]{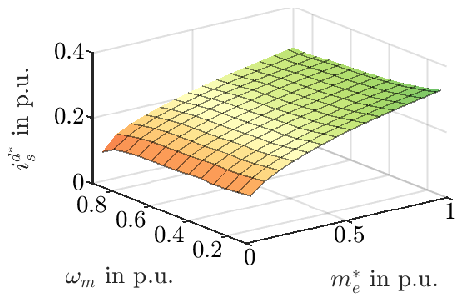}\label{fig:lisd_mept}}
			\subfloat[$\mathcal{L}_{\isqref}^{\mathrm{MEPT}}(\melref, \omegam)$]{\includegraphics[width=4.5cm]{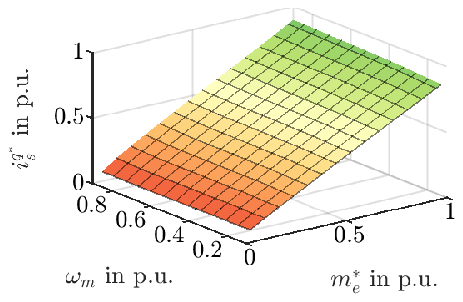}\label{fig:lisq_mept}}
			\caption{Recorded current reference look-up tables (motor mode) for different control strategies.}
			\label{fig:LUTs}
		\end{figure*}
		\subsubsection{Constant V/Hz (M=VHz)}
    		In the open-loop V/Hz control strategy, the excitation of the machine is kept constant (equality constraint), while the torque varies with the slip. 
    		Given a constant V/Hz ratio, the corresponding LUTs $\mathcal{L}_{\isdref}^{\mathrm{VHz}}(\melref,\omegam)$ and $\mathcal{L}_{\isqref}^{\mathrm{VHz}}(\melref,\omegam)$ can be obtained (see Fig.~\ref{fig:lisd_vhz}~\&~\ref{fig:lisq_vhz}). Typically, the nameplate of the machine indicates the rated voltage and frequency, e.g. for rated voltage \SI{400}{\volt} (RMS, line-to-line) and frequency \SI{50}{\hertz}, the rated ratio is $\xi_\idxrated =  \SI{6.53}{\volt\second}$.  
    		It can be shown that, using a different $\xi$, a better (optimized) result in terms of efficiency can be obtained for rated torque (see Fig.~\ref{fig:efficiency_comparison_2D_eta_over_mm}). It should be noted  that if $\xi>\xi_\idxrated$ the machine cannot reach rated speed under full load. 
	
		\subsubsection{Constant flux (M=CF)}
		 In the range below rated speed, the constant flux (CF) strategy prescribes a constant excitation current reference $\isdref = \mathrm{const.}$  (often chosen as the no-load current at rated speed), whereas $\isqref$ is used to control the machine torque. The corresponding LUTs $\mathcal{L}_{\isdref}^{\mathrm{CF}}(\melref,\omegam)$ and $\mathcal{L}_{\isqref}^{\mathrm{CF}}(\melref,\omegam)$ (Figs.~\ref{fig:lisd_cf}~\&~\ref{fig:lisq_cf}) are obtained by evaluating the equality constraint for $\isdref$. 
	
		\subsubsection{Maximum torque per current / ampere (M=MTPC)}
		In order to minimize the ohmic losses of the machine (in fact, only in the stator resistance $\Rs$), MTPC control may be preferred over CF control. Here, evaluating the optimization problem of a minimum stator current magnitude for a given torque reference yields the LUTs $\mathcal{L}_{\isdref}^{\mathrm{MTPC}}(\mel,\omegam)$ and $\mathcal{L}_{\isqref}^{\mathrm{MTPC}}(\melref,\omegam)$ (see Figs.~\ref{fig:lisd_mtpc}~\&~\ref{fig:lisq_mtpc}).
	
    	\subsubsection{Maximum efficiency per torque (MEPT)}
    	As an extension to MTPC, the MEPT control strategy not only reduces stator losses, but also rotor and iron losses. For the LUT generation, the optimization goal of maximizing the efficiency is evaluated. Since the resulting sample data does not run on a smooth curve, fitting the raw data to the function $\isdref = a\cdot\arctan{(b\cdot\melref)}$ (with fitting parameters $a$ and $b$) improves the results. The fitted curves are stored in the LUTs $\mathcal{L}_{\isdref}^{\mathrm{MEPT}}(\melref, \omegam)$ and $\mathcal{L}_{\isqref}^{\mathrm{MEPT}}(\melref, \omegam)$ (see Figs.~\ref{fig:lisd_mept}~\&~\ref{fig:lisq_mept}).

	\section{Discussion of results and efficiency analysis}
	\label{sec:efficiency_analysis}
	The measured efficiencies of the introduced torque control strategies can now be compared for different speeds and torque values. Fig.~\ref{fig:efficiency_comparison_2D_contour} shows 2D contour plots of the machine efficiency for first quadrant operation. Additionally, the respective torque control strategy trajectories  are plotted over the contour plots. Each plot shows results for one specific mechanical speed value, which is kept constant by the load machine. The MEPT and MTPC trajectories are shown for both, measured torque (argument $\mel$) and estimated torque (argument $\melhat$), while for the respective MEPT curves also the sample data points used in the curve fitting are plotted as x's.
	
	It is first observed that the efficiency increases with the speed, which is expected for IMs. However, while the MEPT($\mel$) and MEPT($\melhat$) trajectories drift apart for increasing speeds, the opposite is the case for the respective MTPC($\mel$) and MTPC($\melhat$). More specifically, the MEPT($\melhat$) curve is almost independent of the speed, whereas both MTPC curves and the MTPC($\mel$) curve move to the left. This is easily explained by the fact that the speed dependent core losses are not considered in the efficiency calculation. Interestingly, for the MTPC case, the effect seems to be covered implicitly, although the estimated torque differs from the actual torque. As a consequence, MTPC($\melhat$) might be the better option if no torque sensor is available. Only for very high speeds  ($\omegam > \num{0.9}\,\mathrm{p.u.}$), a difference between MTPC($\melhat$) and MEPT($\mel$) becomes visible.
	
	 Fig.~\ref{fig:efficiency_comparison_2D_eta_over_mm} shows the efficiencies of the different control strategies plotted versus the machine torque for different mechanical speeds. The previous observation of the efficiency increasing for higher speeds is confirmed here. Naturally, the MEPT($\mel$) curve marks the upper limit for all curves, while the difference between the FOC strategies is comparably low for $\mel > \num{0.5}\,\mathrm{p.u.}$. It can be observed that the efficiency difference between the estimated and measured torque MEPT and MTPC curves is almost not visible. 
	 For lower loads, constant flux (CF) operation becomes significantly worse than MEPT and MTPC. Another interesting observation is that FOC seems to be particularly beneficial for lower speeds, since here the efficiency difference between standard V/Hz control and FOC is significant ($\Delta\eta>7\%$ for rated torque), regardless of the FOC control strategy. Conversely, with increasing speed, the V/Hz trajectory approaches the MEPT and MTPC curves, e.g. for $\omegam = \num{0.9}\,\mathrm{p.u.}$, the efficiency difference is only about $1.3\%$. Depending on the choice of the V/Hz ratio, there is either a high performance drop for lower (optimized), or for higher loads (standard); optimizing the V/Hz ratio for the whole operation range is not possible, though.
	 
	 Lastly, Fig.~\ref{fig:LUTs} shows the reference current LUTs for the different control strategies. While $\isqref$ does not vary with $\omegam$ and is almost linear in $\melref$ for all strategies except for V/Hz control, differences are found mainly in the $\isdref$ LUTs: For V/Hz control an almost constant $\isdref$ can be observed (since the excitation is supposed to be constant), for CF control $\isdref$ is perfectly constant, and for MTPC and MEPT control a nonlinear relation in $\melref$ \emph{and} $\omegam$ direction, is observed.

\section{Conclusion}

	An experimental machine identification method, which is based on generic machine equations and steady-state measurements, has been presented. The nonlinear machine maps cover the relevant $dq$-operation range and are obtained without explicit modeling of nonlinear effects, such as magnetic saturation or iron losses. The inferred MEPT control strategy guarantees operation at the maximum achievable efficiency. Furthermore, different torque control strategies have been assessed in terms of efficiency, showing that FOC outperforms V/Hz control in the low-speed range, whereas the difference becomes smaller for higher speeds. It was shown that by chosing a different V/Hz ratio than the rated one, the efficiency curve can be shifted such that better performance is achieved for rated conditions. Unless compensated for, the drawback of speed variations due to slip still persists for V/Hz control, though. Furthermore, it was shown that the MEPT and MTPC curves are both speed-dependent, and that MTPC is the preferred option if no torque sensor is available, since the resulting trajectory matches well with the actual MEPT curve.

\begin{figure*}
		\centering
		\begin{tikzpicture}[every node/.style={font=\footnotesize\bfseries, inner sep=0}]
			\draw[very thick, draw=plca3, dashed] (0,0) -- (0.5,0) node[right, xshift=1mm, anchor=west] (vhz_rated) {V/Hz (rated)};
			\draw[very thick, draw=plca2, dash dot] (vhz_rated.east) ++ (0.25,0) --  ++ (0.5,0) node[right, xshift=1mm, anchor=west] (vhz_opot) {V/Hz (opt.)};		
			\draw[very thick, draw=plca1, dotted] (vhz_opot.east) ++ (0.25,0) --  ++ (0.5,0) node[right, xshift=1mm, anchor=west] (constant_flux) {Constant Flux};		
			\draw[very thick, draw=plca4] (constant_flux.east) ++ (0.25,0) --  ++ (0.5,0) node[right, xshift=1mm, anchor=west] (mtpa) {MTPC ($\mel$)};		
			\draw[very thick, draw=plca4b,dashed] (mtpa.east) ++ (0.25,0) --  ++ (0.5,0) node[right, xshift=1mm, anchor=west] (mtpa2) {MTPC ($\melhat$)};	
			\draw[very thick, draw=plca5] (mtpa2.east) ++ (0.25,0) --  ++ (0.5,0) node[right, xshift=1mm,anchor=west] (mept) {MEPT ($\mel$)};	
			\draw[very thick, draw=plca5b,dashed] (mept.east) ++ (0.25,0) --  ++ (0.5,0) node[right,xshift=1mm, anchor=west] (mept2) {MEPT ($\melhat$)};
			\draw[] (-0.2,-0.3) rectangle ($(mept2.north east) + (0.3,0.2)$);
\end{tikzpicture}
\vspace{0cm}
	
		\centering
		\subfloat[$\omegam = \num{0.3}\,\mathrm{p.u.}$]{\includegraphics[width=8.5cm]{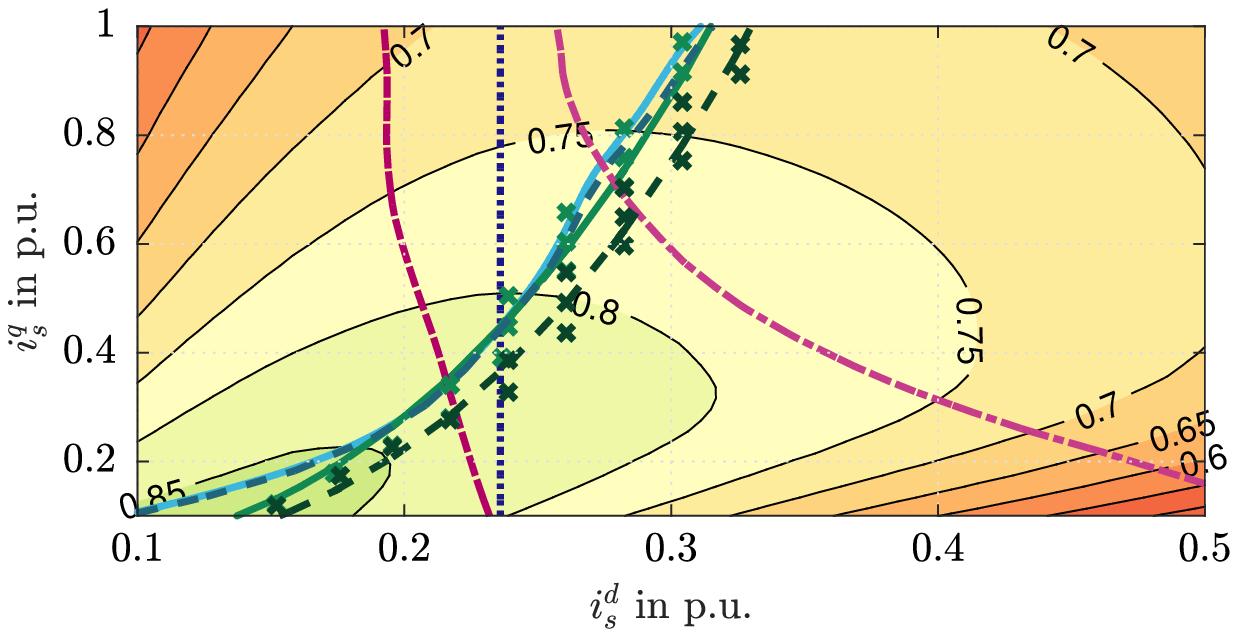}}\quad
		\subfloat[$\omegam = \num{0.5}\,\mathrm{p.u.}$]{\includegraphics[width=8.5cm]{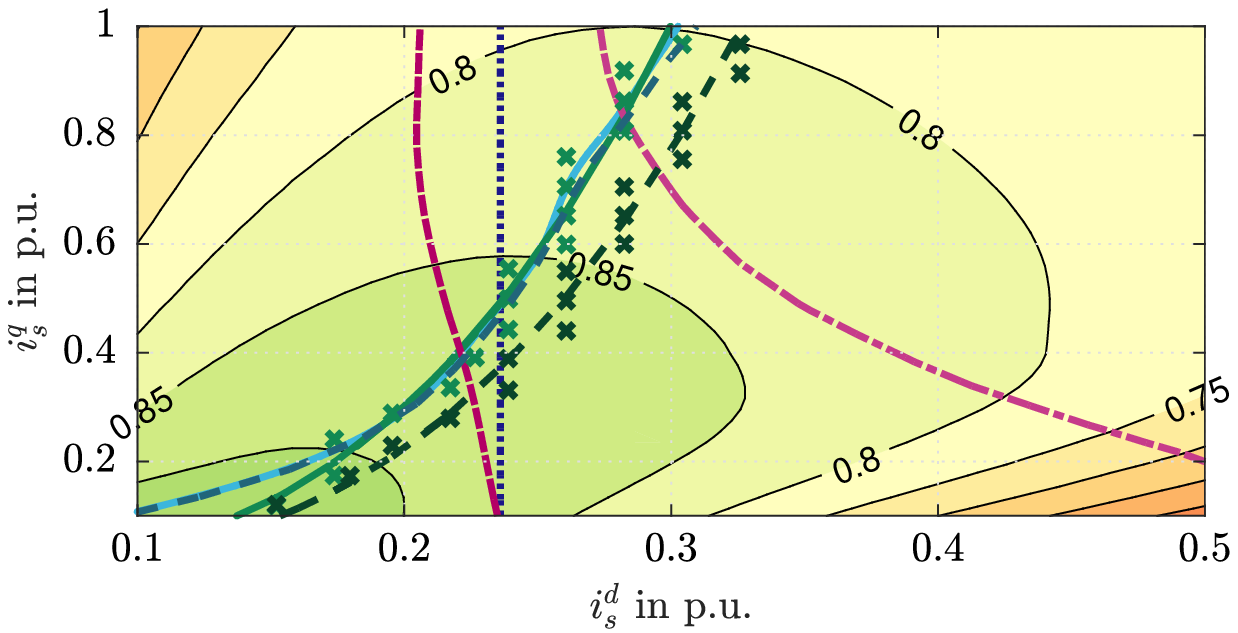}}\\
		\subfloat[$\omegam = \num{0.7}\,\mathrm{p.u.}$]{\includegraphics[width=8.5cm]{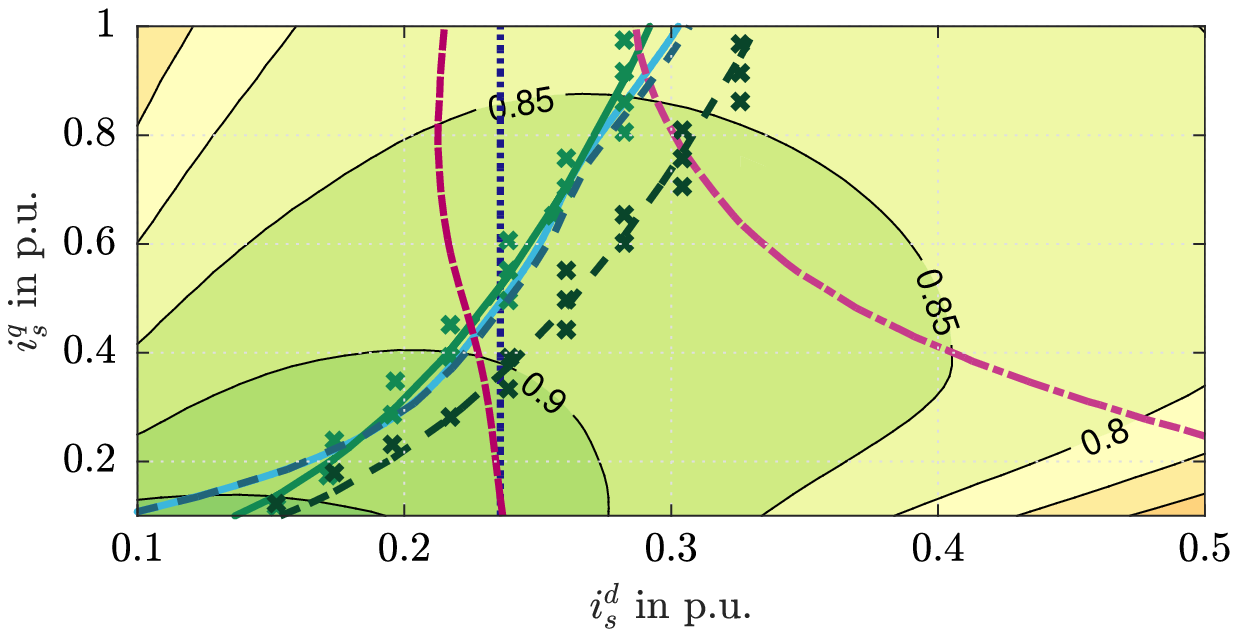}}\quad
		\subfloat[$\omegam = \num{0.9}\,\mathrm{p.u.}$]{\includegraphics[width=8.5cm]{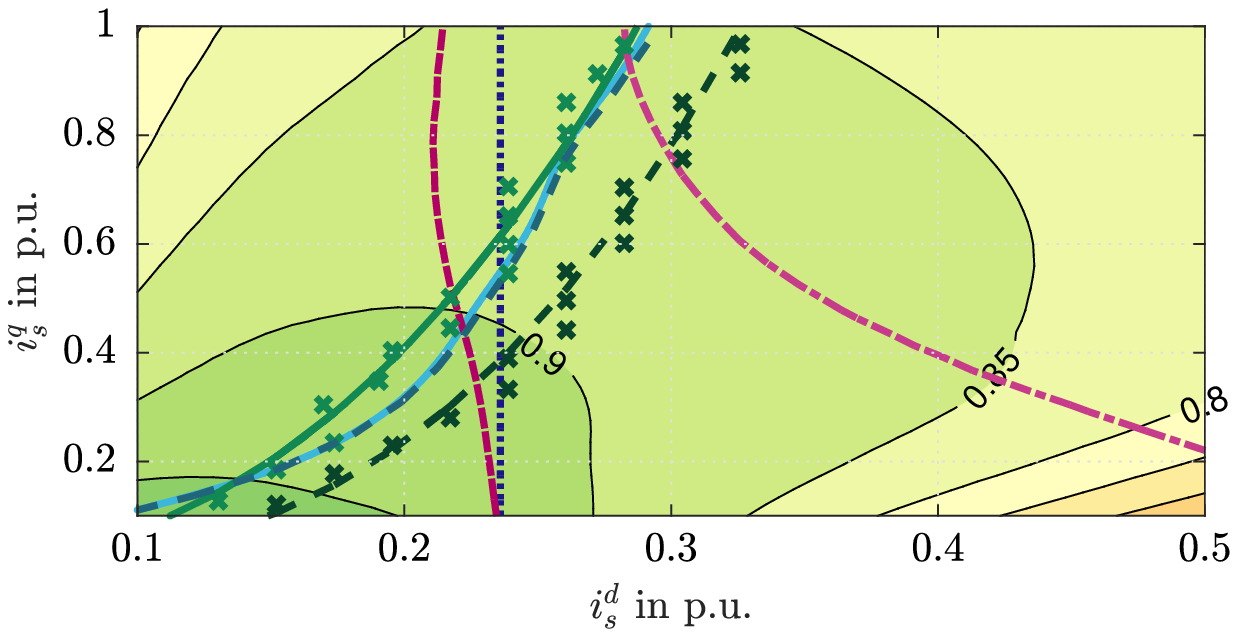}}
		\caption{Efficiency contour plots (first quadrant): Comparison of the torque control strategies for different speeds.}	
		\label{fig:efficiency_comparison_2D_contour}
	\end{figure*}

	\begin{figure*}
		\centering
		\begin{tikzpicture}[every node/.style={font=\footnotesize\bfseries, inner sep=0}]
			\draw[very thick, draw=plca3, dashed] (0,0) -- (0.5,0) node[right, xshift=1mm, anchor=west] (vhz_rated) {V/Hz (rated)};
			\draw[very thick, draw=plca2, dash dot] (vhz_rated.east) ++ (0.25,0) --  ++ (0.5,0) node[right, xshift=1mm, anchor=west] (vhz_opot) {V/Hz (opt.)};		
			\draw[very thick, draw=plca1, dotted] (vhz_opot.east) ++ (0.25,0) --  ++ (0.5,0) node[right, xshift=1mm, anchor=west] (constant_flux) {Constant Flux};		
			\draw[very thick, draw=plca4] (constant_flux.east) ++ (0.25,0) --  ++ (0.5,0) node[right, xshift=1mm, anchor=west] (mtpa) {MTPC ($\mel$)};		
			\draw[very thick, draw=plca4b,dashed] (mtpa.east) ++ (0.25,0) --  ++ (0.5,0) node[right, xshift=1mm, anchor=west] (mtpa2) {MTPC ($\melhat$)};	
			\draw[very thick, draw=plca5] (mtpa2.east) ++ (0.25,0) --  ++ (0.5,0) node[right, xshift=1mm,anchor=west] (mept) {MEPT ($\mel$)};	
			\draw[very thick, draw=plca5b,dashed] (mept.east) ++ (0.25,0) --  ++ (0.5,0) node[right,xshift=1mm, anchor=west] (mept2) {MEPT ($\melhat$)};
			\draw[] (-0.2,-0.3) rectangle ($(mept2.north east) + (0.3,0.2)$);
\end{tikzpicture}
\vspace{0cm}
		\subfloat[$\omegam = \num{0.3}\,\mathrm{p.u.}$]{
			\tikz{
				\node at (0,0) {\includegraphics[width=8.5cm]{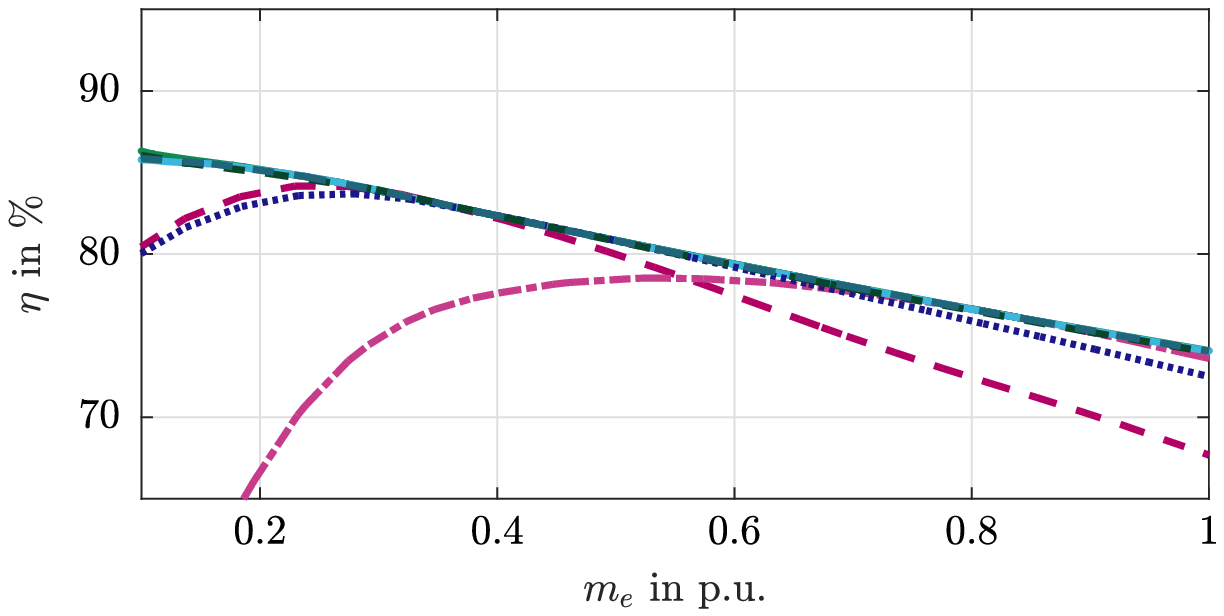}};
				\draw[| - |, red, very thick] (4,-0.25) -- coordinate (a) (4,-1);
				\draw[stealth-, red, thick] (a) ++ (-0.1,0) -- ++ (130:1.5) node[above] {$\Delta\eta = \SI{7.25}{\percent}$};
			}
		}\,
		\subfloat[$\omegam = \num{0.5}\,\mathrm{p.u.}$]{
			\tikz{
				\node at (0,0) {\includegraphics[width=8.5cm]{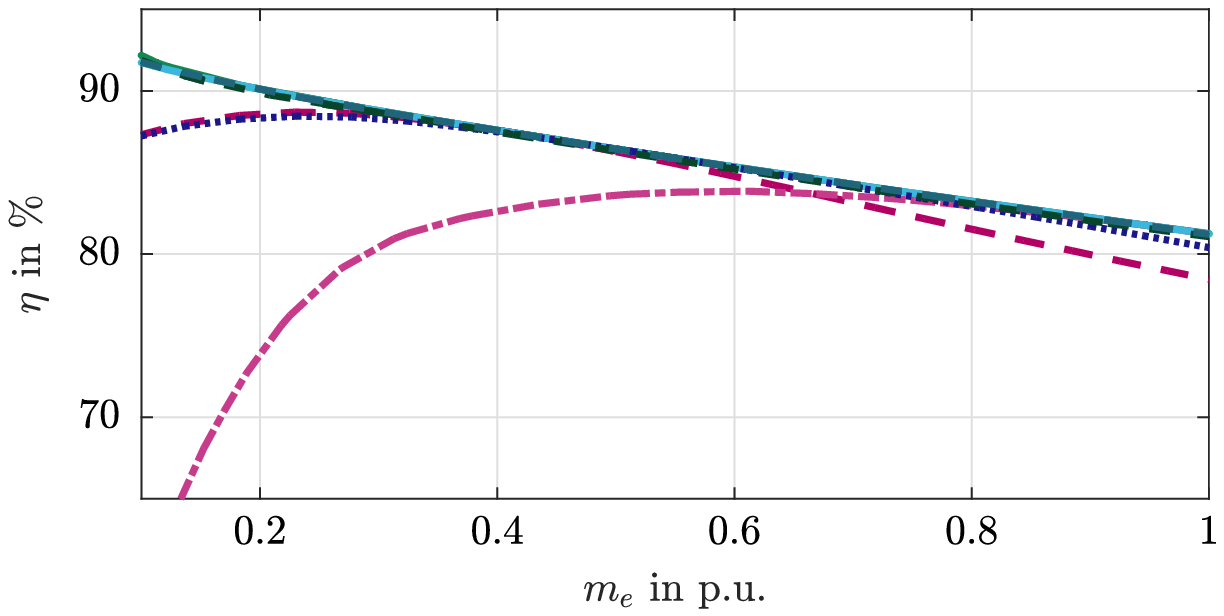}};
				\draw[| - |, red, very thick] (4,0.15) -- coordinate (a) (4,0.53);
				\draw[stealth-, red, thick] (a) ++ (-0.1,0) -- ++ (210:1.3) node[below] {$\Delta\eta = \SI{3.25}{\percent}$};
			}
		}\\
		\subfloat[$\omegam = \num{0.7}\,\mathrm{p.u.}$]{
			\tikz{
				\node at (0,0) {\includegraphics[width=8.5cm]{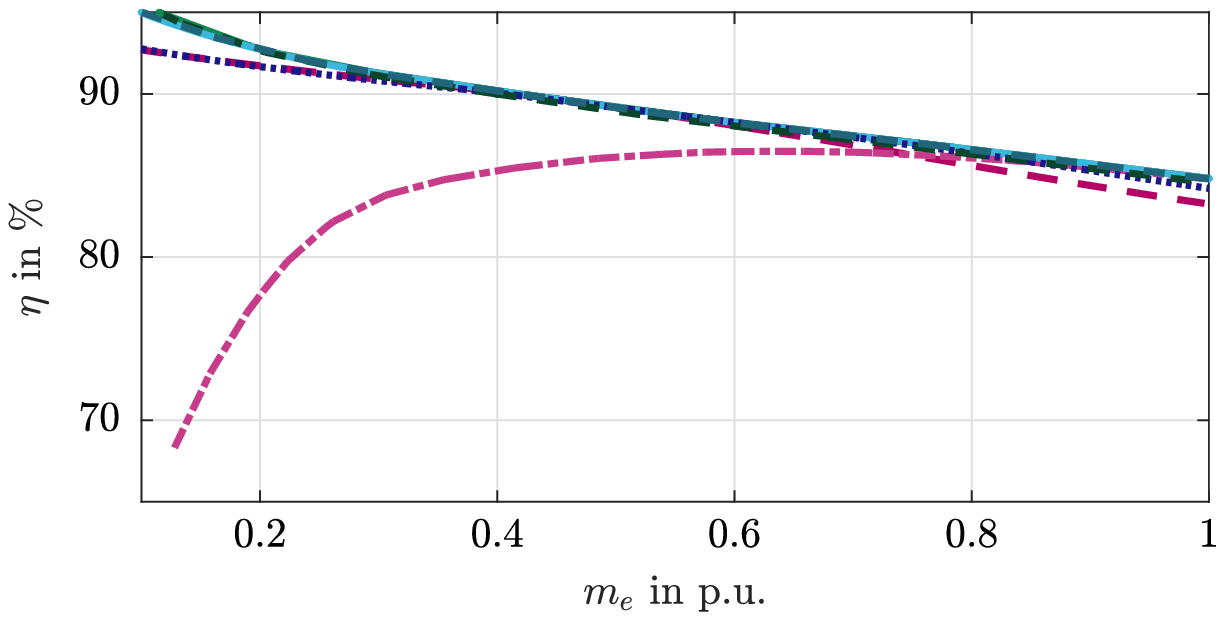}};
				\draw[| - |, red, very thick] (4,0.7) -- coordinate (a) (4,0.9);
				\draw[stealth-, red, thick] (a) ++ (-0.1,0) -- ++ (210:1.3) node[below] {$\Delta\eta = \SI{1.85}{\percent}$};
			}
		}\,
		\subfloat[$\omegam = \num{0.9}\,\mathrm{p.u.}$]{
			\tikz{
				\node at (0,0) {\includegraphics[width=8.5cm]{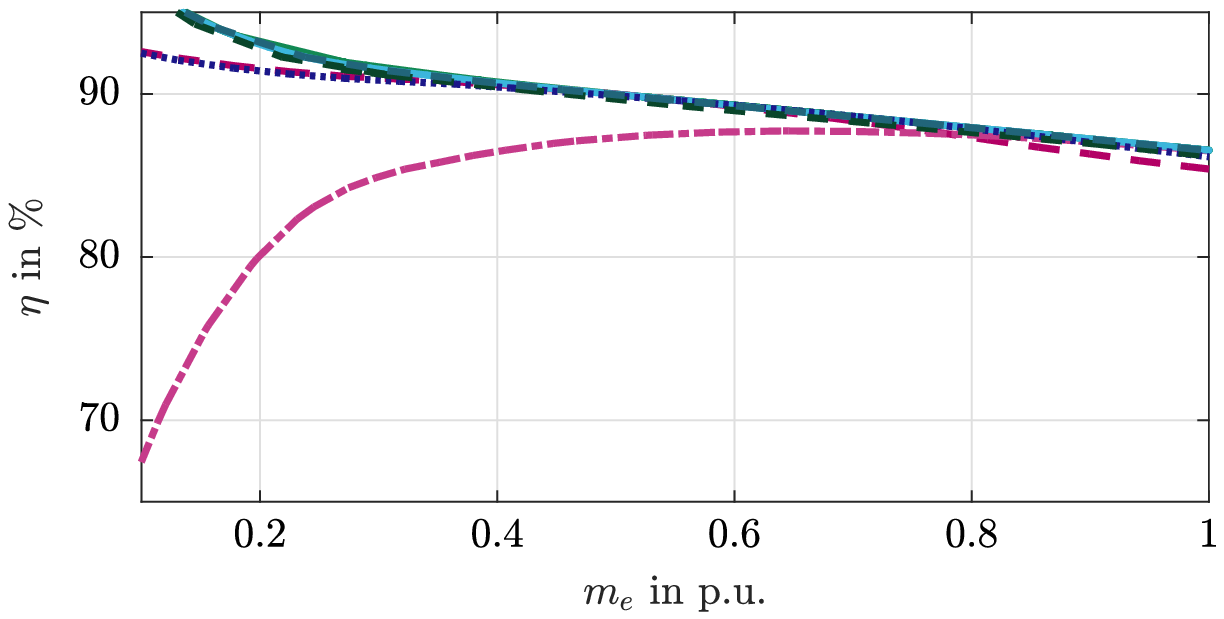}};
				\draw[| - |, red, very thick] (4,.9) -- coordinate (a) (4,1.1);
				\draw[stealth-, red, thick] (a) ++ (-0.1,0) -- ++ (210:1.3) node[below] {$\Delta\eta = \SI{1.3}{\percent}$};
			}
		}
		\caption{Efficiency over load torque: Comparison of the different torque control strategies for different speeds.}	
		\label{fig:efficiency_comparison_2D_eta_over_mm}
	\end{figure*}

\bibliographystyle{Bibliography/IEEEtranTIE}
\bibliography{Bibliography/IEEEabrv,Bibliography/lib}\ 

\end{document}